\documentclass[twocolumn]{aastex63}

\graphicspath{{./}{figures/}}

\usepackage{color}
\usepackage{amsmath}
\usepackage{xcolor}
\usepackage{mathrsfs}
\usepackage{natbib}
\usepackage{gensymb}
\usepackage{xspace}
\usepackage{subfigure}

\newcommand{\Rj}{\ensuremath{R_{\rm{Jup}}}\xspace}
\newcommand{\Mj}{\ensuremath{M_{\rm{Jup}}}\xspace}

\newcommand{\Msun}{\ensuremath{M_{\odot}}\xspace}
\newcommand{\angstrom}{\textup{\AA}}

\newcommand{\caltech}{Department of Astronomy, California Institute of Technology, Pasadena, CA 91125, USA}
\newcommand{\gps}{Division of Geological \& Planetary Sciences, California Institute of Technology, Pasadena, CA 91125, USA}
\newcommand{\ucsc}{Department of Astronomy \& Astrophysics, University of California, Santa Cruz, CA95064, USA}
\newcommand{\keck}{W. M. Keck Observatory, 65-1120 Mamalahoa Hwy, Kamuela, HI, USA}
\newcommand{\ucla}{Department of Physics \& Astronomy, 430 Portola Plaza, University of California, Los Angeles, CA 90095, USA}
\newcommand{\jpl}{Jet Propulsion Laboratory, California Institute of Technology, 4800 Oak Grove Dr.,Pasadena, CA 91109, USA}
\newcommand{\ucsd}{Center for Astrophysics and Space Sciences, University of California, San Diego, La Jolla, CA 92093}

\newcommand{\berkeley}{Department of Astronomy, University of California at Berkeley, CA 94720, USA}

\newcommand{\osu}{Department of Astronomy, The Ohio State University, 100 W 18th Ave, Columbus, OH 43210 USA}
\newcommand{\ifahilo}{Institute for Astronomy, University of Hawaii at Hilo, 640 N Aohoku Pl, Hilo, HI 96720, USA}

\begin{document}

\title{A Clear View of a Cloudy Brown Dwarf Companion from High-Resolution Spectroscopy}

\correspondingauthor{J. Xuan}
\email{jxuan@astro.caltech.edu}

\author[0000-0002-6618-1137]{Jerry W. Xuan}
\affiliation{\caltech}

\author[0000-0003-0774-6502]{Jason Wang}
\altaffiliation{51 Pegasi b Fellow}
\affiliation{\caltech}
\affiliation{Center for Interdisciplinary Exploration and Research in Astrophysics (CIERA) and Department of Physics and Astronomy, Northwestern University, Evanston, IL 60208, USA}

\author[0000-0003-2233-4821]{Jean-Baptiste Ruffio}
\affiliation{\caltech}

\author[0000-0002-5375-4725]{Heather Knutson}
\affiliation{\gps}

\author{Dimitri Mawet}
\affiliation{\caltech}
\affiliation{\jpl}

\author{Paul Mollière}
\affiliation{Max-Planck-Institut für Astronomie, Königstuhl 17, 69117 Heidelberg, Germany}

\author{Jared Kolecki}
\affiliation{\osu}

\author{Arthur Vigan}
\affiliation{Aix Marseille Univ, CNRS, CNES, LAM, Marseille, France}

\author{Sagnick Mukherjee}
\affiliation{Department of Astronomy and Astrophysics, University of California, Santa Cruz, CA 95064, USA}

\author{Nicole Wallack}
\affiliation{\gps}

\author[0000-0002-4361-8885]{Ji Wang}
\affiliation{\osu}

\author{Ashley Baker}
\affiliation{\caltech}
\altaffiliation{51 Pegasi b Fellow}

\author{Randall Bartos}
\affiliation{\jpl}

\author{Geoffrey A. Blake}
\affiliation{\gps}

\author{Charlotte Z. Bond}
\affiliation{UK Astronomy Technology Centre, Royal Observatory, Edinburgh EH9 3HJ, United Kingdom}

\author{Marta Bryan}
\affiliation{\berkeley}
\altaffiliation{51 Pegasi b Fellow}

\author{Benjamin Calvin}
\affiliation{\caltech}
\affiliation{\ucla}

\author{Sylvain Cetre}
\affiliation{\keck}

\author{Mark Chun}
\affiliation{\ifahilo}

\author[0000-0001-8953-1008]{Jacques-Robert Delorme}
\affiliation{\keck}
\affiliation{\caltech}

\author{Greg Doppmann}
\affiliation{\keck}

\author{Daniel Echeverri}
\affiliation{\caltech}

\author[0000-0002-1392-0768]{Luke Finnerty}
\affiliation{\ucla}

\author[0000-0002-0176-8973]{Michael P. Fitzgerald}
\affiliation{\ucla}

\author{Katelyn Horstman}
\affiliation{\caltech}

\author{Julie Inglis}
\affiliation{\gps}

\author[0000-0001-5213-6207]{Nemanja Jovanovic}
\affiliation{\caltech}

\author{Ronald L\'opez}
\affiliation{\ucla}

\author[0000-0002-0618-5128]{Emily C. Martin}
\affiliation{\ucsc}

\author{Evan Morris}
\affiliation{\ucsc}

\author{Jacklyn Pezzato}
\affiliation{\caltech}

\author{Sam Ragland}
\affiliation{\keck}

\author{Bin Ren}
\affiliation{\caltech}

\author[0000-0003-4769-1665]{Garreth Ruane}
\affiliation{\caltech}
\affiliation{\jpl}

\author{Ben Sappey}
\affiliation{\ucsd}

\author{Tobias Schofield}
\affiliation{\caltech}

\author{Andrew Skemer}
\affiliation{\ucsc}

\author{Taylor Venenciano}
\affiliation{Physics and Astronomy Department, Pomona College, 333 N. College Way, Claremont, CA 91711, USA}

\author[0000-0001-5299-6899]{J. Kent Wallace}
\affiliation{\jpl}

\author{Peter Wizinowich}
\affiliation{\keck}

\begin{abstract}

Direct imaging studies have mainly used low-resolution spectroscopy ($R\sim20-100$) to study the atmospheres of giant exoplanets and brown dwarf companions, but the presence of clouds has often led to degeneracies in the retrieved atmospheric abundances (e.g. C/O, metallicity). This precludes clear insights into the formation mechanisms of these companions. The Keck Planet Imager and Characterizer (KPIC) uses adaptive optics and single-mode fibers to transport light into NIRSPEC ($R\sim35,000$ in $K$ band), and aims to address these challenges with high-resolution spectroscopy. Using an atmospheric retrieval framework based on \texttt{petitRADTRANS}, we analyze KPIC high-resolution spectrum ($2.29-2.49~\mu$m) and archival low-resolution spectrum ($1-2.2~\mu$m) of the benchmark brown dwarf HD~4747~B ($m=67.2\pm1.8~\Mj$, $a=10.0\pm0.2$ au, $T_{\rm eff}\approx1400$~K). We find that our measured C/O and metallicity for the companion from the KPIC high-resolution spectrum agree with that of its host star within $1-2\sigma$. The retrieved parameters from the $K$ band high-resolution spectrum are also independent of our choice of cloud model. In contrast, the retrieved parameters from the low-resolution spectrum are highly sensitive to our chosen cloud model. Finally, we detect CO, H$_2$O, and CH$_4$ (volume mixing ratio of log(CH$_4$)=$-4.82\pm0.23$) in this L/T transition companion with the KPIC data. The relative molecular abundances allow us to constrain the degree of chemical disequilibrium in the atmosphere of HD~4747~B, and infer a vertical diffusion coefficient that is at the upper limit predicted from mixing length theory.

\end{abstract}


\section{Introduction} \label{sec:intro}

The Keck Planet Imager and Characterizer (KPIC) is a new suite of instrument upgrades at Keck II, including a single-mode fiber injection unit \citep{mawet_Observing_2017, delorme_Keck_2021a} that feeds light into the upgraded NIRSPEC \citep{martin_overview_2018, lopez_Characterization_2020}, enabling high-resolution spectroscopy (HRS\footnote{We will use HRS to abbreviate both high-resolution spectroscopy (the technique) and high-resolution spectra (the data) in this paper. The same is true for LRS: low-resolution spectra or spectroscopy.}) at $R\sim35,000$ in $K$ band. By using single-mode fibers to inject light from planets and brown dwarfs at high-contrast, KPIC provides suppression of the stellar point-spread function at the fiber input and a stable line spread function that is independent of incoming wavefront aberrations \citep{mawet_Observing_2017, wang_Detection_2021}. By observing at high-resolution, we can further distinguish between star and planet light from their spectral differences \citep{wang_Observing_2017,mawet_Observing_2017}. Recently, \citet{wang_Detection_2021} published the KPIC detections of HR~8799~c,~d,~e, demonstrating the ability of KPIC to detect molecular lines and measure the rotational line broadening of planets at high contrast ($\Delta K\approx11$) and small separations ($\approx0.4$") from their host star. 

The atmospheric composition of a substellar object holds a wealth of information about its formation, accretion, and evolutionary history, as well as fundamental physical processes that shape its atmosphere. It is therefore important to assess how well KPIC and other fiber-fed, high-resolution spectrographs (e.g. Subaru/REACH \citealt{kotani_Extremely_2020} and VLT/HiRISE \citealt{otten_Direct_2021}) can measure the atmospheric compositions of directly imaged planets and brown dwarfs. Specifically, previous studies of gas giant planet formation have highlighted the carbon-to-oxygen ratio (C/O) and metallicity (e.g. [C/H]) of the atmosphere as informative probes of formation history \citep[e.g.][]{oberg_effects_2011, madhusudhan_c/o_2012, piso_Snowline_2015}. To first order, a companion with a C/O and metallicity similar to that of its host star is consistent with formation via gravitational instability in a protostellar disk or fragmentation in a molecular cloud, akin to binary star formation  \citep{bate_formation_2002}. On the other hand, differences between the companion and stellar C/O are suggestive of core accretion \citep{pollack_formation_1996} as the likely formation mechanism, and in that scenario, could be used to constrain where the companion formed in the disk relative to ice lines of major C- and O-bearing molecules (e.g. H$_2$O, CO$_2$, and CO). This picture can be complicated by a variety of effects such as the relative amount of solids incorporated into the planet's atmosphere \citep[e.g.][]{madhusudhan_Chemical_2014, oberg_Excess_2016, mordasini_Imprint_2016, nowak_Peering_2020, pelletier_Where_2021}.

So far, atmospheric characterization of directly imaged companions has mostly relied on low-resolution spectroscopy (LRS) with resolving powers of $R\approx20-100$. LRS is sensitive to continuum emission originating from the deepest observable layer of the atmosphere and modified by opacity sources further up. Many of these companions have temperatures warm enough for silicate clouds to condense in their atmospheres \citep{marley_Cool_2015}, and there is much evidence that cloud opacity plays an important role in the LRS of directly imaged companions and brown dwarfs with L or L/T transition spectral types \citep[e.g.][]{skemer_Directly_2014, burningham_retrieval_2017, nowak_Peering_2020}. However, due to our limited knowledge of cloud physics, a reliable assessment of atmospheric abundances from LRS could be fraught with degeneracies between clouds, the pressure-temperature profile, and chemical abundances \citep[e.g.][]{burningham_retrieval_2017}. In addition, the retrieval results can also be highly sensitive to systematics in different data sets that are combined to obtain a wider wavelength coverage \citep{wang_Chemical_2020a}. More encouragingly, \citet{molliere_Retrieving_2020} report atmospheric abundances that are relatively robust to clouds and model choices, though \citet{burningham_Cloud_2021} show that issues such as an unphysically small radius could persist despite improvements in cloud modeling and extensive wavelength coverage ($1-15\mu$m).

Recently, \citet{wang_Retrieving_2022} presented the first atmospheric free retrievals at high-resolution for a directly imaged companion. They studied the L-type brown dwarf HR~7672~B ($T_{\rm eff}\approx1800$~K) using KPIC HRS and near-infrared photometry, and measured carbon and oxygen abundances that are consistent within $<1.5\sigma$ to that of its host star. In this paper, we present a detailed atmospheric study of HD~4747~B using both KPIC HRS ($K$ band) and archival low-resolution spectra (LRS) from 1-2.2~$\mu$m that we re-extract in a uniform manner. While the KPIC HRS resolves individual molecular lines and conains direct information about a companion's atmospheric abundances, LRS provides spectral shape and luminosity measurements, which has the potential to complement the HRS.

Compared to HR~7672~B, HD~4747~B is a colder L/T transition object ($T_{\rm eff} \approx1400$~K) with strong evidence for clouds and a similar color to directly imaged planets such as HR 8799 c,d,e \citep{crepp_GPI_2018, peretti_orbital_2019}. Like HR~7672~B, the wealth of prior knowledge available for HD~4747~B makes it a valuable benchmark object to test whether we can make robust inferences with spectroscopic data. First, we are able to precisely measure the dynamical mass of HD~4747~B (\S~\ref{sec:orbit}). Mass is a fundamental quantity that is poorly constrained for most directly imaged companions \citep{bowler_imaging_2016}. Furthermore, given its high mass, HD~4747~B is expected to have formed via direct gravitational collapse in the same cloud or disk as its host star, which means that we can assume chemical homogeneity: the brown dwarf and primary star should share the same chemical composition. Finally, with the companion mass, observed luminosity, and stellar age, we can independently estimate the brown dwarf's radius from evolutionary models.

In this paper, we use the open-source radiative transfer code \texttt{petitRADTRANS} \citep{molliere_petitRADTRANS_2019, molliere_Retrieving_2020} to fit the HRS and LRS for HD~4747~B in a retrieval framework. The main goals of our study are to measure the atmospheric composition of this brown dwarf companion using both the HRS and LRS, and to present a detailed characterization of its atmosphere, including constraints on clouds, chemical equilibrium or disequilibrium, and the detection of CH$_4$. In this process, we also explore the relative advantages and disadvantages of HRS versus LRS.

This paper is organized as follows: in \S~\ref{sec:system_prop}, we summarize the system properties including our mass measurement for HD~4747~B. Our spectroscopic data and data reduction procedure is described in \S~\ref{sec:all_data}. We then discuss our spectral analysis framework in \S~\ref{sec:model_framework}. We present individual and joint retrievals of the HRS and LRS in \S~\ref{sec:hrs_results}, \S~\ref{sec:lrs_results} and \S~\ref{sec:joint_results}, respectively. We summarize the lessons learned in \S~\ref{sec:discuss}, and conclude in \S~\ref{sec:conclude}.

\section{System properties} \label{sec:system_prop}
\subsection{Host star} \label{sec:host}
In this section, we summarize relevant properties of the host star. HD~4747 is a main-sequence, solar-type star located $\approx19$ parsec away based on its Gaia eDR3 parallax \citep{brown_Gaia_2021}. Chromospheric emission in the Ca II H\&K lines are visible in the stellar spectrum (log$R_{\rm HK}=-4.72\pm0.02$), which \citet{peretti_orbital_2019} used to derive an age of $2.3\pm1.4$ Gyr from the age-log$R_{\rm HK}$ calibration of \citet{mamajek_Improved_2008}. This agrees with the gyro-chronological age estimate of $3.3^{+2.3}_{-1.9}$ Gyr from \citet{crepp_GPI_2018}. These studies also converged on $T_{\rm eff}$ around 5300–5400 K, and a surface gravity log($g$) of 4.5–4.65. Of particular relevance to this study are the C/O ratio and metallicity of the host star, since we expect these to be roughly similar to those of the brown dwarf. HD~4747 is found to have a sub-solar metallicity, with [Fe/H]=$-0.23\pm0.05$ from \citet{peretti_orbital_2019} and [Fe/H]=$-0.22\pm0.04$ from \citet{crepp_GPI_2018}. Previous studies including  \citet{brewer_SPECTRAL_2016} and \citet{peretti_orbital_2019} also measured the elemental abundances for the star, but either did not take into account non-local thermodynamic equilibrium (LTE) effects on their oxygen abundances \citep{amarsi_Carbon_2019} or do not quote error bars.  We instead carry out a new analysis using the method described in \citet{kolecki_Searching_2021} to derive the abundances for different elements, and correct the results to account for 3D non-LTE effects \citep{amarsi_Carbon_2019} on the results. For this analysis, we used an archival spectrum from FEROS \citep{kaufer_FEROS_1997} which covers 350-920 nm at $R=48,000$. Using this spectrum, we measure the equivalent widths of absorption lines and compare them to model stellar atmospheres in an iterative approach using the MOOG code \citep{sneden_nitrogen_1973}. From our derived carbon and oxygen abundances, we find C/O=$0.48\pm0.08$. The iron abundance is [Fe/H]=$-0.30\pm0.5$, in agreement with previous studies. 

Since Fe condenses out for temperatures below $\approx1800$~K \citep{marley_Cool_2015}, it is not a relevant gaseous absorber in the photosphere of HD~4747~B. Therefore, the more useful metrics for comparison are C and O. From our analysis above, we find [C/H]$=-0.08\pm0.06$ and [O/H]$=-0.02\pm0.04$ for the host star. [C/H] is defined as $\log_{10} ({N_{\rm C}/N_{\rm H}})_{\rm star}$ - $\log_{10} ({N_{\rm C}/N_{\rm H}})_{\rm sun}$, where $N_{\rm C}$ and $N_{\rm H}$ are the number fraction of C and H respectively. [O/H] is defined similarly. We adopt \citet{asplund_Chemical_2009} as our solar reference in order to be consistent with \texttt{petitRADTRANS}, which we use to model the atmosphere of HD~4747~B. 

\subsection{Orbit and Dynamical mass} \label{sec:orbit}
\begin{deluxetable}{lc}
\tablecaption{Selected parameters from orbit fit} \label{tab:orbit}
\tablehead{\colhead{Parameter} & \colhead{Value}} 

\startdata
$M$ ($\Msun$) & $0.85\pm0.04$ \\
$m$ ($\Mj$) & $67.2\pm1.8$ \\
$a$ (AU) & $10.0\pm0.2$ \\
inclination (deg) & $48.0\pm0.9$ \\
ascending node (deg) & $89.4\pm1.1$ \\
period (yr) & $33.2\pm0.4$ \\
argument of periastron (deg) & $267.2\pm0.5$ \\
eccentricity & $0.7317\pm0.0014$ \\
Epoch of periastron (JD) & $2462615\pm155$ \\
\enddata
\tablecomments {The dynamical mass of the host star, which is fit as a free parameter, agrees well with isochrone-derived masses from \citet{peretti_orbital_2019} and \citet{crepp_GPI_2018}.}
\end{deluxetable}

The orbit and mass of HD~4747~B have been measured by several studies using relative astrometry from Keck/NIRC2, host star radial velocities (RV) from Keck/HIRES, and Gaia-Hipparcos absolute astrometry \citep{brandt_precise_2019, xuan_Evidence_2020}. Here, we take advantage of 23 yr of RV observations published in \citet{rosenthal_California_2021} and the improved precision of the Hipparcos-Gaia Catalog of Accelerations (HGCA) \citep{brandt_hgca_2021a} based on Gaia eDR3 \citep{brown_Gaia_2021} to update the orbit and mass of HD~4747~B. HD~4747~B shows significant proper motion anomalies (PMa) in both the Gaia and Hipparcos epochs, with S/N of $77.2$ and $9.1$, and the position angle and amplitude of the PMa is consistent with being induced by the brown dwarf companion. For the relative astrometry, we use data points tabulated in \citet{brandt_precise_2019}, except for the two GPI epochs measured by \citet{crepp_GPI_2018}, which we replaced with our new measurements from \S~\ref{sec:lrs_data}. We choose not to use the companion RV as measured by KPIC for this fit, because it does not appreciably improve our already well-constrained orbital solution. 

To fit the relative astrometry, radial velocity, and absolute astrometry from Gaia and Hipparcos together, we use the \texttt{orvara} package \citep{brandt_orvara_2021}, which is designed to jointly fit these types of data and takes into account the Gaia and Hipparcos astrometry at the epoch astrometry level using \texttt{htof} \citep{brandt_htof_2021}. We use the priors listed in Table 4 of \citet{brandt_orvara_2021} for the fitted parameters. The posterior is sampled using the parallel-tempering MCMC sampler \citep{vousden_Dynamic_2016}, a fork of emcee \citep{foreman-mackey_emcee:_2013} over 50000 steps with 10 temperatures and 100 walkers. The fits converged as determined by visual inspection of the chains, and we discarded the first ten percent as burn-in. In \texttt{orvara}, the system parallax and other linear parameters are analytically marginalized out to speed up the fits.

The resulting orbit and mass measurements are tabulated in Table~\ref{tab:orbit}, while the model fits are shown in Appendix A. We find a companion mass of $m=67.2\pm1.8~\Mj$, which is consistent with previous values, but more precise. We checked the \texttt{orvara} results with a second fit where we model the Gaia and Hipparcos astrometry using the methodology in \citet{xuan_Evidence_2020}. This gives $m=67.1\pm2.0~\Mj$, consistent with the \texttt{orvara} result. Furthermore, the companion mass and orbital parameters we find are also consistent with results from orbit fits that only use RV and imaging data (no Gaia-Hipparcos astrometry) from \citet{peretti_orbital_2019} and \citet{crepp_GPI_2018}. We adopt the companion mass from our \texttt{orvara} fit for the spectral analysis in this paper. 

\section{Spectroscopic data} \label{sec:all_data}

\subsection{High-resolution spectroscopy} \label{sec:hrs_data}

\subsubsection{KPIC Observations}
We observed HD~4747~B on UT 2020 September 28 with Keck/NIRSPEC. The data were collected using the first version of the KPIC fiber injection unit (FIU) \citep{delorme_Keck_2021a}. The FIU is located downstream of the Keck II adaptive optics system and is used to inject light from a selected object into one of the single mode fibers connected to NIRSPEC. We obtain spectrum in K band, which is broken up into nine echelle orders from 1.94-2.49~$\mu$m. The observing strategy is identical to that of \citet{wang_Detection_2021}. In short, we placed the companion on the fiber with the highest throughput and acquired six exposures of 600 seconds each, for a total integration time of 1 hour. The relative astrometry of the companion was computed using \href{http://whereistheplanet.com/}{whereistheplanet.com} \citep{wang_where_2021}, based on data in \citet{peretti_orbital_2019}. For calibration purposes, we acquired a pair of 60 second exposures of the host star before observing the companion, and a pair of 60 second exposures of a telluric standard star (HIP 6960) after the companion exposures so as to share nearly the same airmass. Using exposures on the host star, we calculated an end-to-end throughput from the top of the atmosphere to the detector of $1.8-2.0\%$ during the observations.

\begin{figure*}[t!]
\centering
\begin{subfigure}
  \centering
  \includegraphics[width=.4\linewidth]{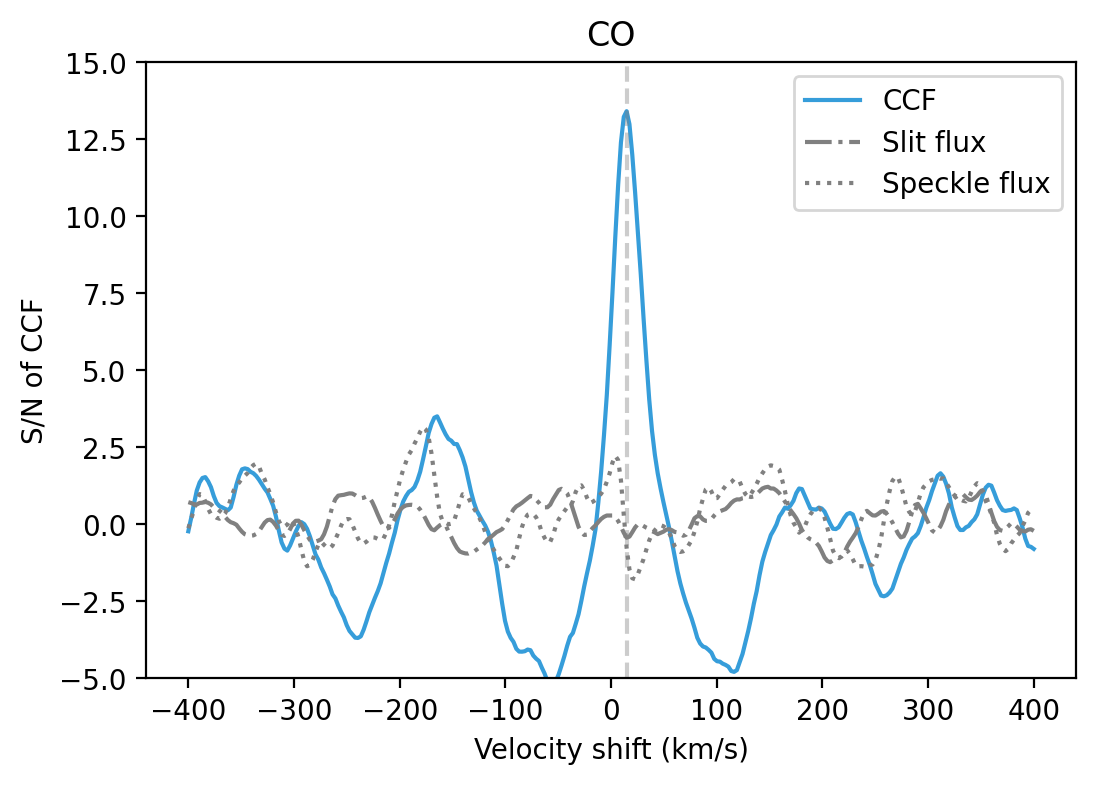}
  \hspace{5mm}
\end{subfigure}
\begin{subfigure}
  \centering
  \includegraphics[width=.4\linewidth]{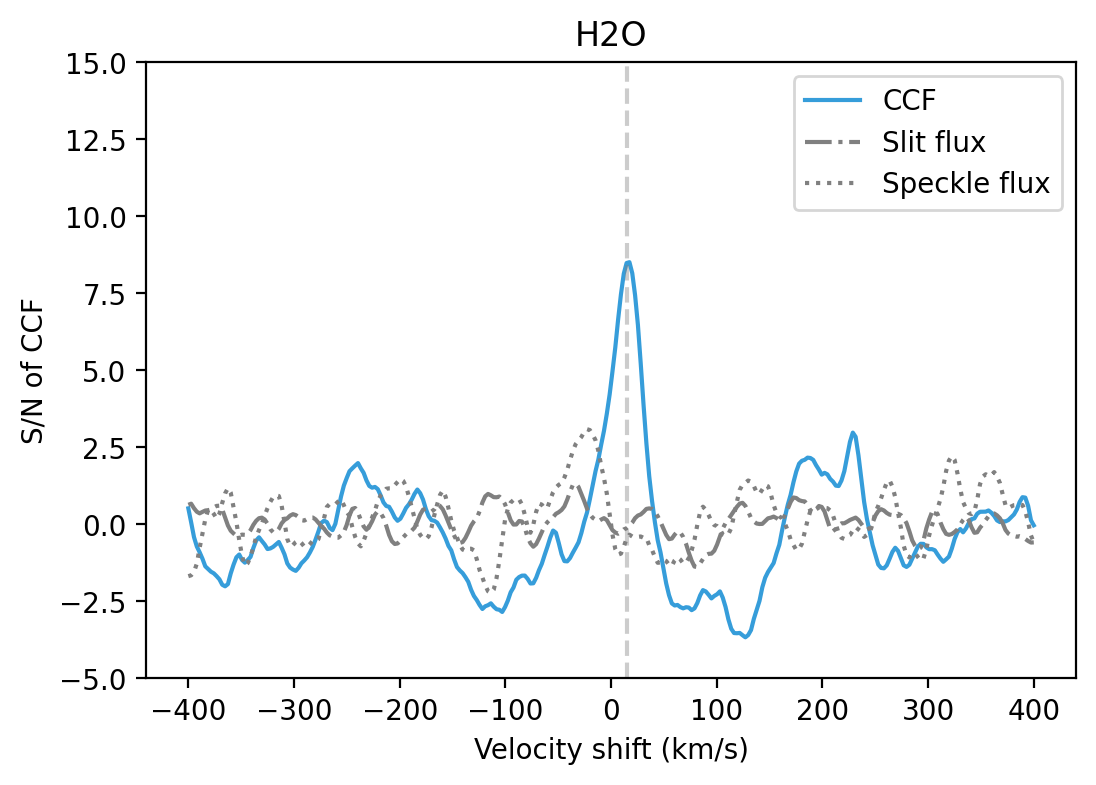}
\end{subfigure}
\caption{Cross-correlation functions (CCFs) in blue show detections of CO and H$_2$O using 3 spectral orders of the KPIC HRS. Gray lines represent CCFs of the background flux (from the slit) and speckle flux, whose standard deviations are used as estimates of the CCF noise. The vertical dashed lines at $15$ km/s show the expected RV of the companion from its known orbit. The strong structure in the blue CCFs outside the peaks arise because we only fit single molecules here.}
\label{fig:ccf_detect}
\end{figure*}

\subsubsection{Data reduction}
To extract the spectra from the raw data, we follow the procedure outlined in \citet{wang_Detection_2021}, which the KPIC team has implemented in a public Python pipeline.\footnote{\url{https://github.com/kpicteam/kpic_pipeline}} The images for all objects were reduced in the same way. First, we removed the thermal background from the images using combined instrument background frames taken during daytime. As shown in \citet{wang_Detection_2021}, the thermal background of our data is dominated by the warm optics rather than the sky background. We also remove persistent bad pixels identified using the background frames. Then, we use data from the telluric standard star to fit the trace of each column in the four fibers and nine spectral orders, which give us the position and standard deviation of the PSF (in spatial direction) at each column. The trace positions and widths were additionally smoothed using a cubic spline in order to mitigate random noise. We adopt the trace locations and widths as the line spread function (LSF) positions and widths in the dispersion dimension.

For every frame, we then extracted the 1D spectra in each column of each order. To remove residual background light, we subtracted the median of pixels that are at least 5 pixels away from every pixel in each column. Finally, we used optimal extraction to sum the flux using weights defined by the 1D Gaussian LSF profiles calculated from spectra of the telluric star.

The extracted spectra have a median signal-to-noise ratio (S/N) of $\approx8$ per pixel element, which has a typical width of $0.2~\angstrom$, and consists of a mixture of light from the brown dwarf companion and stellar speckles. The S/N of KPIC is optimized for wavelengths around $2.3~\mu$m, where CO has a series of strong absorption lines. For our analysis, we use three spectral orders from 2.29-2.49~$\mu$m, which contain the strongest absorption lines from the companion and have relatively few telluric absorption lines. Note that the three spectral orders have gaps in between them, so we have data over $\approx0.13~\mu$m (instead of $0.2~\mu$m; see Fig.~\ref{fig:kpic_mod_data}).

As a preliminary analysis, we cross-correlate our KPIC spectra with single-molecule templates assuming $T_{\rm eff}=1400$~K and log($g$)=5.5 from the Sonora model grid \citep{marley_Sonora_2021}. In short, we estimate the maximum likelihood value for both the single-molecule companion flux and speckle flux in the data as a function of RV (radial velocity) shift using the method described in \citet{wang_Detection_2021}, which is based on \citet{ruffio_thesis_2019}. We find that H$_2$O and CO are detected with S/N of 8.5 and 13.5 respectively (Fig.~\ref{fig:ccf_detect}). CH$_4$ is not detected with statistical confidence in this crude analysis, but we present evidence for a weak CH$_4$ detection in \S~\ref{sec:ch4_detect}.

\subsection{Low-resolution spectroscopy} \label{sec:lrs_data}

\subsubsection{Gemini Planet Imager IFS}
\label{sec:gpi_extract}
The Gemini Planet Imager (GPI) observed HD~4747~B on UT 2015 December 24 and 25, in the $K_1$ (1.90-2.19~$\mu$m, $R=66$) and $H$ (1.50-1.80~$\mu$m, $R=46.5$) bands, respectively, and the data were published in \citet{crepp_GPI_2018}. After doing some fits to the published spectrum, we found that the average flux levels of the $K_1$ and $H$ bands are inconsistent, and the error bars appear to be significantly over-estimated.

We therefore re-extracted the GPI spectrum using the \texttt{pyKLIP} package \citep{wang_pyKLIP_2015}, which models a stellar point spread function (PSF) with Karhunen-Loève Image Processing (KLIP, also known as Principal Component Analysis) following the framework in \citet{soummer_Detection_2012b} and \citet{pueyo_DETECTION_2016}. We tested various model choices to minimize the residuals after stellar PSF subtraction while preserving the companion signal. A key parameter we tuned was the number of Karhunen-Loève (KL) modes. KL modes represent an orthogonal basis for patterns in the images that are used to model the stellar PSF.  We chose 5 and 12 KL modes to subtract the stellar PSF in the $H$ and $K_1$ band data, respectively. After subtracting the stellar PSF, we first extracted the companion's relative astrometry in terms of separation and position angle, which are reported in Appendix B (Table~\ref{tab:gpi_astrom}). Then, we extracted the flux at the companion's determined location as a function of wavelength, which gave us the raw spectrum. Note that rather than using spectral differential imaging (SDI) to subtract the stellar PSF, we only used angular differential imaging (ADI). For a bright companion like HD~4747~B, ADI is more than sufficient to properly remove the PSF of the star given sufficient parallactic angle rotation.

To flux-calibrate the raw spectrum, we used the satellite spot flux ratios\footnote{\href{https://www.gemini.edu/instrumentation/gpi/capability}{https://www.gemini.edu/instrumentation/gpi/capability}} to find the companion-to-star flux ratio. To obtain the observed flux density of the companion, we empirically determined the flux scaling factor $R^2/d^2$ by fitting a PHOENIX model \citep{husser_new_2013} of the star ($T_{\rm eff}=5400$, log($g$)=4.5, and [Fe/H] = -0.5) using the star's 2MASS J, H, K \citep{cutri_2MASS_2003} and the Gaia G band magnitudes \citep{riello_Gaia_2021}. The zeropoint fluxes and filter transmission of the photometric bands are downloaded from the SVO Filter Service\footnote{\href{http://svo2.cab.inta-csic.es/theory/fps/}{http://svo2.cab.inta-csic.es/theory/fps/}} and the Gaia website.\footnote{\href{https://www.cosmos.esa.int/web/gaia/edr3-passbands}{https://www.cosmos.esa.int/web/gaia/edr3-passbands}} To obtain measurement uncertainties, we injected 20 fake companions at the same separation and equally spaced position angles in the data, and repeated the same spectral extraction process. We avoided using the fake injections that were within $20\degree$ of the real companion to avoid biasing the fluxes. We inflated the uncertainties on the extracted spectra by $2.5\%$ to account for errors in the stellar flux calibration. The value of $2.5\%$ is estimated by comparing our empirically computed flux scaling factor with the value of $R^2/d^2$ of the star (using the radius from \citealt{peretti_orbital_2019} and the Gaia parallax).

\subsubsection{SPHERE IFS}
\label{sec:sphere_data}
HD~4747~B was observed on UT 2016 December 12 and 2017 September 28 with the Spectro-Polarimetric High-contrast Exoplanet Research (SPHERE; \citealt{beuzit_SPHERE_2019}). The SPHERE Integral Field Spectrograph (IFS) \citep{claudi_SPHERE_2008} collects data in the $YH$ band from 0.95-1.6~$\mu$m ($R=29$). The extracted spectra was published in \citet{peretti_orbital_2019}, but is not available. We therefore reduced the raw data using the SPHERE pipeline \citep{vigan_vltsphere_2020}, and performed a similar post-processing procedure with pyKLIP as described above for the GPI spectra. The only difference is that we needed to use ADI+SDI to perform PSF-subtraction for the SPHERE IFS data, which did not have enough parallactic angle rotation (only $\approx0.2\degree$). For the SPHERE IFS data, flux calibration is based on unocculted observations of the host star. We chose to use the 2017 data for our analysis since it was taken under much better observing conditions and yields slightly higher spectral S/N than the 2016 data, despite shorter integration times. 

\begin{figure*}[t!]
    \centering
    \includegraphics[width=0.75\linewidth]{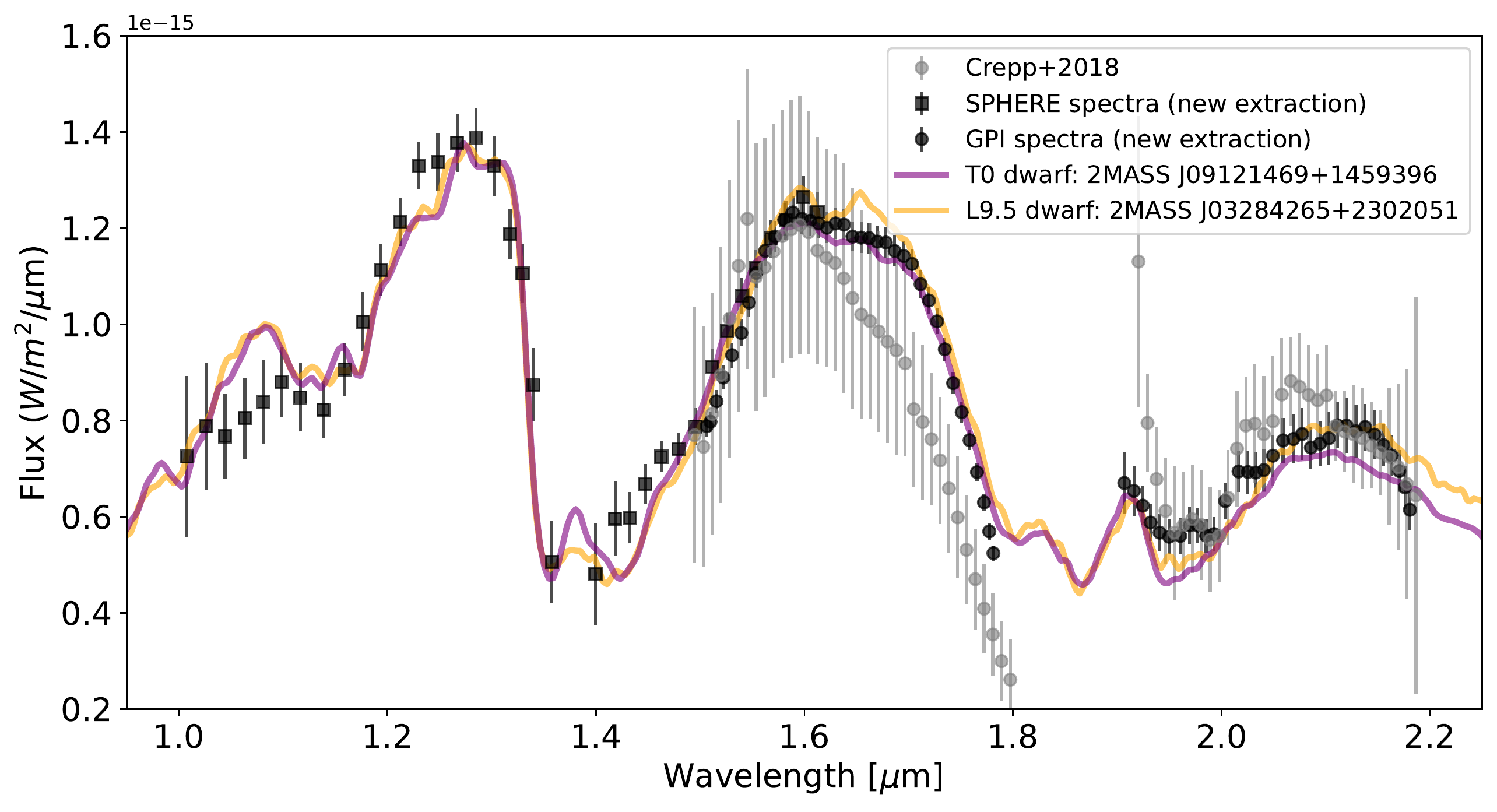}
    \caption{Our extracted LRS of HD~4747~B from GPI and SPHERE are plotted in black circles and squares, respectively, and the published spectra for GPI is shown in gray dots from \citet{crepp_GPI_2018}. Overplotted in color are spectra of a L9.5 dwarf and T0 dwarf from SPEX \citep{burgasser_SpeX_2014}, which show good agreement with our extracted spectra, demonstrating that HD~4747~B is consistent with a spectral type near the L/T transition.}
    \label{fig:gpi_spec}
\end{figure*}

\subsubsection{Results and comparison with previous LRS}
Our newly extracted GPI and SPHERE spectra are plotted in black circles and squares respectively in Fig.~\ref{fig:gpi_spec}, and available in Appendix B (Table~\ref{tab:gpi_spec}). The absolute flux scaling of our GPI spectrum agrees well with the published spectrum in gray from \citet{crepp_GPI_2018}, but the uncertainties are much smaller. The shape of our new SPHERE spectrum also agrees well with that in \citet{peretti_orbital_2019}. From the PSF-subtracted images, the brown dwarf companion is detected with a median S/N per wavelength bin of $\approx61$ and $\approx26$ in the GPI $H$ and $K_1$ bands, and $\approx20$ in the SPHERE data. When comparing the extracted spectrum to that of field brown dwarfs from the SPEX library \citep{burgasser_SpeX_2014} in Fig.~\ref{fig:gpi_spec}, we find that the newly extracted GPI spectrum is in better agreement compared to the previously published spectrum. As in \citet{crepp_GPI_2018} and \citet{peretti_orbital_2019}, we find a spectral type near the L/T transition (the best matching spectra were from a L9.5 and T0 dwarf). The SPHERE IFS spectrum increases our wavelength coverage by a factor of $\sim2$, which we find is important for constraining model atmosphere parameters in our fits to the LRS.

\section{Spectral analysis} \label{sec:model_framework}

\subsection{Forward modeling the KPIC high-resolution spectrum}
Here, we briefly describe the framework to forward model and fit the HRS from KPIC, which follows \citet{wang_Detection_2021}. When a companion of interest is aligned with one of the KPIC fibers, the companion light and a fraction of light from the host star's speckle field are injected into the fiber. At the projected separation of HD~4747~B ($\approx0.6$\arcsec), we find the speckles are roughly the same brightness as our companion ($K_s\approx14.4$ from \citealt{crepp_TRENDS_2016}). Furthermore, the light is transmitted through Earth's atmosphere and modulated by the instrument optics. Thus, we build two forward models (one for the companion, one for speckles) and jointly fit them as a linear combination. Below we detail how we generate each of the model components.

The companion spectral templates are generated with \texttt{petitRADTRANS}. We shift the templates in wavelength space to fit for the radial velocity. Then, we rotationally broaden the templates by a projected rotation rate $v\sin{i}$ using the \texttt{fastRotBroad} function in \texttt{PyAstronomy} \citep{Czesla_pyA_2019}, and convolve the templates with the instrumental LSF. The effect of limb darkening is included in \texttt{petitRADTRANS} by integrating intensities along multiple angles between the ray and atmospheric normal.

Next, we multiply the companion model by the telluric response function, which characterizes the atmospheric transmission as a function of wavelength and includes telluric absorption lines. The telluric model is calculated by dividing the spectrum of the standard star (HIP 6960) by a PHOENIX stellar model with matching properties ($T_{\rm eff}=9200$ and log($g$)=4.0). 

To model the speckle contribution to the data, we use on-axis observations of the host star taken before the companion exposures. These observations are reduced in the same way as the companion spectra, but have much higher S/N. Unlike the companion models, the host star observations are already modulated by telluric transmission.

The last step is to remove continuum variations. The KPIC spectra are not flux-calibrated and contain a smoothly varying continuum due to stellar speckles and wavelength-dependent atmospheric refraction. Therefore, we apply high-pass filtering with a median filter of 100 pixels ($\sim0.002~\mu$m) on both the data and models to subtract the continuum following \citet{wang_Detection_2021}. To determine the optimal filter size, we carried out a series of injection-recovery tests, and found that $\sim100$ pixels is best at recovering weak molecular signals for our data set. Larger filter sizes (e.g. 200 pixels or more) do not remove the continuum sufficiently, and smaller filter sizes (50 pixels or less) tend to be overly aggressive at removing weak molecular signals. 

Finally, we flux-normalize both the companion and stellar models and multiply them by different flux scaling factors, which are fitted parameters. The flux scales are in units of counts as measured by the NIRSPEC detector. After scaling, the companion and speckle models are added and the same high-pass filter is applied on the final model before fitting it to the data.

\subsection{Atmospheric retrieval setup} \label{sec:retr_setup}
We implement a `retrieval' framework based on \texttt{petitRADTRANS} to model the data, which means that we freely retrieve the chemical abundances, vertical temperature structure, and cloud properties from the data. Previous studies have used retrievals to model HRS of self-luminous exoplanets and brown dwarfs \citep[e.g.][]{burningham_retrieval_2017,molliere_Retrieving_2020}, and show that it can be a powerful alternative to fitting self-consistent grid models, which solve for the abundances and temperature profiles with physical assumptions such as chemical equilibrium. The retrieval approach allows more flexibility to fit the data and can potentially provide much more detailed information about the atmospheric properties, with the caveat that it is important to check for physical plausibility of the models since retrievals need not be self-consistent. 

In our main set of retrievals, we fit for the chemical abundances in terms of C/O and atmospheric metallicity [C/H]\footnote{We denote the atmospheric metallicity as [C/H] because we are only sensitive to C- and O-bearing molecules in this brown dwarf's atmosphere.} along with a quench pressure (where the chemical timescale of a certain reaction is equal to the mixing timescale) to allow for disequilibrium chemistry, the temperature profile (\S~\ref{sec:temp_chem}), the cloud structure (\S~\ref{sec:cloud_choice}), and other parameters such as the radius. We denote these quenched chemistry retrievals to distinguish from free retrievals where the abundances of each gas species is fit independently. Each component of the model is described in the subsections below. 
We use the correlated-k and line-by-line opacity sampling methods in \texttt{petitRADTRANS} for the low-resolution and high-resolution retrievals respectively. For high-resolution, we include opacities for CO, H$_2$O, CH$_4$, NH$_3$, and CO$_2$, and for low-resolution we additionally include Na and K. This is because the alkali lines have wings which affect the $\sim1~\mu$m portion of the LRS, while their opacities are negligible over the portion of $K$ band covered by our HRS. We repeated our baseline HRS retrieval with Na and K included and found that the addition of these two species did not influence the results or improve the fit.

Because the native high-resolution opacities are at $R=10^6$, much higher than the resolution of our HRS resolution ($R\approx35,000$), we down-sampled the opacity tables by a factor of six in order to speed up the retrievals (by roughly the same factor) and reduce the corresponding computational cost. We checked that the maximum deviation in synthetic spectra obtained by using the down-sampled opacities relative to the full-resolution opacity model is $<5\%$ of the minimum HRS error bars. In addition, we repeated our fiducial HRS retrieval with the native opacities ($R=10^6$) and found that it yielded the same results. We re-binned the correlated-k opacities to $R=200$ for our fits to the LRS, which has a maximum resolution of $66$. We also repeated our fiducial LRS retrieval at the native $R=1000$ opacities and found the results are fully consistent. 
 
\subsubsection{Temperature structure and chemistry} \label{sec:temp_chem}
We retrieve the pressure-temperature (PT) profile of the brown dwarf between $P=10^{-4} - 10^{3}$ bars, which sets the vertical extent of the atmosphere. We use the P-T profile parametrization from \citet{molliere_Retrieving_2020} which has six free parameters. The spatial coordinate is an optical depth $\tau = \delta P^{\alpha}$, where $\delta$ and $\alpha$ are the first two parameters. The atmosphere then consists of a high altitude region (top of atmosphere to $\tau = 0.1$) fitted with three temperature points equi-distant in log pressure, a middle radiative region ($\tau = 0.1$ to radiative-convective boundary) which uses the Eddington approximation with $T_0$ as the `internal temperature', and a lower region (radiative-convective boundary to bottom of atmosphere), which is set to follow the moist adiabatic temperature gradient once the atmosphere becomes unstable to convection \citep{molliere_Retrieving_2020}. We ignore stellar irradiation as a source of heat because the total incident energy on HD~4747~B at periastron ($\approx2.7$~au) is approximately four orders of magnitude less than its luminosity, which is dominated by the brown dwarf's internal energy. 

In our quenched chemistry retrievals, the C/O, [C/H], and P-T profile determine the equilibrium chemical abundances (mass fractions of molecules) as a function of pressure, by interpolating the chemical equilibrium table from \citet{molliere_Retrieving_2020}. The opacities we include in the models are listed in \S~\ref{sec:retr_setup}. In \texttt{petitRADTRANS}, the abundances of all metals except oxygen are assumed to scale together such that [C/H] = [Si/H] = [N/H], etc. Then, C/O and [C/H] are combined to set the oxygen abundance \citep{molliere_Retrieving_2020}. We use \citet{asplund_Chemical_2009} as our reference for the solar metallicity in these models.

Finally, we include a quench pressure $P_{\rm quench}$ which fixes the abundances of H$_2$O, CO, and CH$_4$ where $P < P_{\rm quench}$ using the equilibrium values found at $P_{\rm quench}$ \citep{Zahnle_methane_2014, molliere_Retrieving_2020}. The inclusion of $P_{\rm quench}$ allows for the possibility of disequilibrium chemistry, which occurs where the atmospheric mixing timescale is shorter than the chemical reaction timescale. We only include a quench pressure for the net reaction between H$_2$O, CO, and CH$_4$ because these molecules are the only ones detectable in our KPIC HRS (see \S~\ref{sec:ch4_detect} for the CH$_4$ detection), and chemical kinetics modeling indicates that the abundances of these three molecules are closely linked to each other by a series of reactions \citep[e.g.][]{moses_Chemical_2013}. In summary, our quenched chemistry retrievals use C/O, [C/H] and $P_{\rm quench}$ to set the abundances of each gas species for a given P-T profile.

\subsubsection{Clouds} \label{sec:cloud_choice}
\citet{crepp_GPI_2018} and \citet{peretti_orbital_2019} analyzed LRS for HD~4747~B and found evidence for a cloudy atmosphere. We summarize their results in Table~\ref{table:bayes_factors} along with our new measurements. In this study, we consider both clear and cloudy models in order to explore the sensitivity of our retrieved abundances to the assumed cloud properties. For our cloudy model, we use the EddySed model from \citet{ackerman_Precipitating_2001} as implemented in \texttt{petitRADTRANS} \citep{molliere_Retrieving_2020}. In this model, the cloud particles both absorb and scatter the outgoing photons from the atmosphere according to measured optical properties \citep{molliere_petitRADTRANS_2019}. The cloud particles can be either crystalline or amorphous, and the opacities of the clouds are computed assuming either homogeneous and spherical particles, modeled with Mie theory, or irregularly-shaped cloud particles, modeled with the Distribution of Hollow Spheres (DHS) \citep{min_Modeling_2005, molliere_petitRADTRANS_2019}.

For HD~4747~B, we consider models with two different cloud species (MgSiO$_{3}$ and Fe) and properties (amorphous or crystalline particles). We choose to focus on MgSiO$_{3}$ and Fe for several reasons. First, the condensation curves of these two species intersect the thermal profile of a $T_{\rm eff}=1400$~K, log($g$)=5.5 object from the Sonora atmospheric model \citep{marley_Sonora_2021} at $\sim10-50$ bars. While the Sonora model is cloudless, it provides a rough estimate of which cloud species are relevant. Second, recent theoretical work has shown that MgSiO$_{3}$ is expected to be the most important cloud species for substellar objects with $T_{\rm eff}>950$~K due to its low nucleation energy barriers and the relatively high elemental abundances of Mg, Si, and O \citep{gao_Aerosol_2020}. Finally, studies using mid-IR spectroscopy from Spitzer have found direct evidence for a MgSiO$_{3}$ absorption feature at $\sim10~\mu$m in field brown dwarfs \citep{cushing_Spitzer_2006,luna_Empirically_2021}, and specifically amorphous MgSiO$_{3}$ \citep{burningham_Cloud_2021}. Although MgSiO$_{3}$ and Fe clouds do not have distinct features in the near-IR, they still impact the near-IR spectrum by contributing a wavelength-dependent opacity. Our baseline model uses amorphous MgSiO$_{3}$ modeled with Mie theory (abbreviated MgSiO$_{3}$, `am') for the clouds. In addition, we also consider models with MgSiO$_{3}$, `cd', which assumes crystalline cloud particles modeled with DHS, as well as models with two cloud species (MgSiO$_{3}$ + Fe) for the LRS. 

\begin{figure*}[t!]
    \centering
    \includegraphics[width=1.0\linewidth]{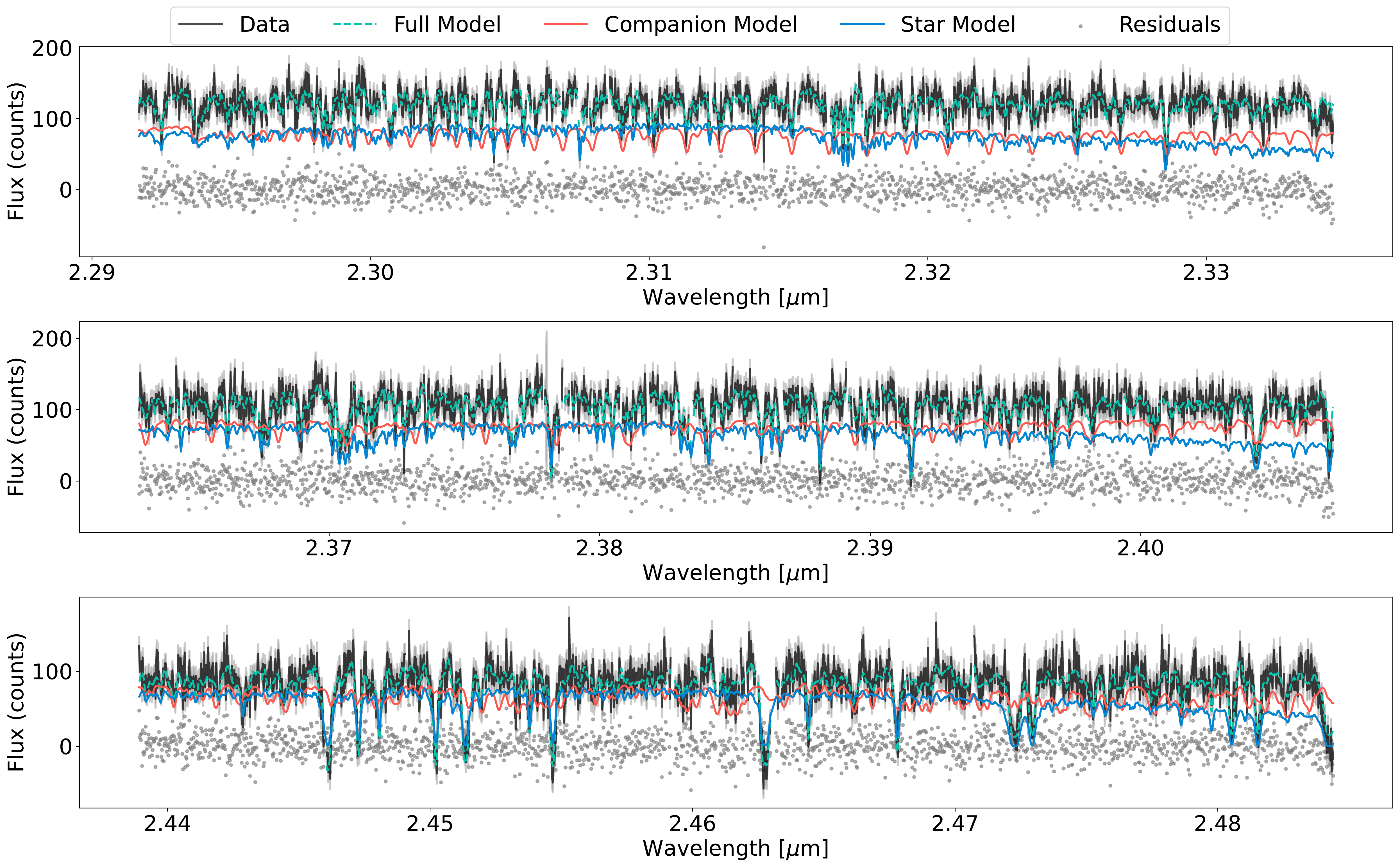}
    \caption{The KPIC HRS used in this study are plotted in black, with error bars inflated to the best fit value in gray. A sample full model is shown in teal (dashed), and consists of the companion model in orange (which has been RV shifted and broadened), and the stellar model in blue to model the speckle contribution. The companion model shown does not include tellurics to focus on molecular features, but tellurics are included in our fits. The residuals are shown as gray points.}
    \label{fig:kpic_mod_data}
\end{figure*}

\subsubsection{Methane opacities} \label{sec:ch4_opa}
Given that HD~4747~B is located near the L/T transition for brown dwarfs, we might expect to observe methane in its atmosphere. Previous $L$ band studies have detected methane in field brown dwarfs with spectral types as early as mid-L, or up to $T_{\rm eff}\approx1800$~K \citep{noll_Onset_2000a, johnston_NIRSPEC_2019}. In this study, we adopted the HITEMP CH$_4$ line list from \citet{hargreaves_Accurate_2020}, which we convert into opacities following the \texttt{petitRADTRANS} documentation. When cross-correlating a model generated with the HITEMP CH$_4$ opacities with a late T dwarf, we obtained a CCF S/N of $\approx15$, in comparison to $\approx5$ when cross-correlating with a model generated from the default CH$_4$ opacities from ExoMol \citep{yurchenko_ExoMol_2014} in \texttt{petitRADTRANS}. 

\subsubsection{Additional fit parameters}
\texttt{petitRADTRANS} computes the flux density as emitted at the surface of the object. For the LRS, we scale the model by the distance and companion radius, where the radius is another free parameter, and the distance is taken from the Gaia eDR3 parallax \citep{brown_Gaia_2021}. For the HRS, we also fit the companion's radial velocity and $v\sin{i}$, as well as an error multiple term to account for any underestimation in the data uncertainties.

Due to imperfect starlight subtraction in the spectral extraction process, we found that our LRS likely still contains correlated noise from the wavelength-dependence of speckles, as has been noted by several previous studies on high-contrast companions \citep[e.g.][]{derosa_Spectroscopic_2016, samland_Spectral_2017a, currie_SCExAO_2018, wang_Keck_2020, wang_Constraining_2021}. This is evident in the residual frames, where we can see speckles at 5-20\% of the companion intensity in the PSF-subtracted images. We therefore adopt a Gaussian process with a squared exponential kernel to empirically estimate the correlated noise in the GPI H, K and SPHERE YJH bands when fitting models to the data. Following \citet{wang_Keck_2020}, we assume that our extracted error bars contain a fraction $f_{\rm amp}$ of correlated noise, and $1-f_{\rm amp}$ of white noise, and fit for $f_{\rm amp}$ and the scale of correlation $l$. This adds $2 \times 3 = 6$ additional parameters to the retrievals. 

As an alternative model, we also tried fitting the LRS with error inflation terms and flux scaling factors for the SPHERE and GPI spectra along the lines of \citet{molliere_Retrieving_2020}, but found that our results were very sensitive to our choice of prior for the flux scaling factor. We conclude that our GP model is better suited to account for correlated noise from speckles, and use it in all LRS fits presented in this work.

\subsection{Priors}\label{sec:priors}
We adopt uniform or log-uniform priors for all model parameters except for the mass, for which we use a Gaussian prior of $67.2\pm1.8~\Mj$ from the dynamical mass measurement (\S~\ref{sec:orbit}). For the parametric P-T profile parameters, we exclude profiles that contain temperature inversions, as the heat budgets of widely separated companions are dominated by their internal luminosities. For the companion's radius, we use a uniform prior between 0.6 - 1.2 $\Rj$. When including a quench pressure, we use a log-uniform prior from $10^{-4}$ - $10^{3}$, which is the full pressure range of our models. The priors for all retrieval parameters are tabulated in Appendix C.

\subsubsection{Model fitting with nested sampling} \label{sec:ns_bayes}
We use nested sampling as implemented by \texttt{dynesty} \citep{speagle_DYNESTY_2020} to find the posterior distributions for the model parameters. Specifically, we use 200 live points and adopt the stopping criterion that the estimated contribution of the remaining prior volume to the total evidence is less than 1\%. We repeated a few retrievals using 1000 live points and found the evidence remains roughly the same, implying the fits have converged when using 200 live points.

One advantage of adopting nested sampling is that we can use the Bayesian evidence from each fit to calculate the Bayes factor $B$, which assesses the relative probability of model $M_2$ compared to $M_1$. We will use the Bayes factor to compare different models throughout this paper to determine whether a given $M_2$ is justified over $M_1$. In Table~\ref{table:bayes_factors}, we take a baseline model (MgSiO$_3$, am) to be $M_1$ and compare other models to it. Based on \citet{jeffrey_1983}, a model with 100 times lower $B$ than the model with the highest $B$ can be `decisively' rejected. $B$ of $\lesssim10$ is considered weak evidence for preferring one model over the other. We first run retrievals with only the HRS (\S~\ref{sec:hrs_results}), only the LRS (\S~\ref{sec:lrs_results}), as well as joint retrievals with both HRS and LRS (\S~\ref{sec:joint_results}). 

\begin{deluxetable*}{cc|ccccc|c}
\tablecaption{Spectral retrievals carried out on HD~4747~B. The right most column lists the Bayes factor ($B$) for each retrieval, with the EddySed (MgSiO$_3$, am) model as the baseline model with $B=1$ (see \S~\ref{sec:ns_bayes} for an explanation of model comparison with $B$). We adopt the first row (in bold) as our final results for this paper. A few key parameters and their central 68\% credible interval with equal probability above and below the median are listed, along with values for common parameters from two previous studies. For our cloudy models, `am' stands for amorphous cloud particles + Mie scattering, and `cd' stands for crystalline particles + DHS model, as described in \S~\ref{sec:cloud_choice}. Except for the HRS model labeled chemical equilibrium, all our other models are quenched chemistry retrievals. \label{table:bayes_factors}}
\tabletypesize{\footnotesize}
\tablehead{
Data/Reference & Cloud Model & C/O & [C/H] & Radius (\Rj) & log($g$) & $T_\textrm{eff}$ (K) & $B$
}
\startdata
\textbf{HRS (KPIC)} & \textbf{EddySed (MgSiO$_3$, am)} & $\mathbf{0.66\pm0.04}$ & $\mathbf{-0.10^{+0.18}_{-0.15}}$ & $\mathbf{0.82^{+0.19}_{-0.13}}$ & $\mathbf{5.39 ^{+0.15}_{-0.18}}$ & $\mathbf{1652^{+128}_{-218}}$ & 1.0 \\
HRS & EddySed (MgSiO$_3$, cd) & $0.67\pm0.04$ & $-0.06^{+0.23}_{-0.18}$ & $0.90\pm0.19$ & $5.32^{+0.20}_{-0.17}$ & $1577^{+167}_{-253}$ & 1.15 \\
HRS & Clear & $0.67^{+0.05}_{-0.04}$ & $-0.09^{+0.24}_{-0.16}$ & $0.87^{+0.19}_{-0.17}$ & $5.34^{+0.19}_{-0.17}$ & $1677^{+132}_{-142}$ & 0.61 \\
HRS & Clear (chemical equilibrium) & $0.60\pm0.02$ & $0.73^{+0.40}_{-0.31}$ & $0.69^{+0.12}_{-0.06}$ & $5.27^{+0.20}_{-0.14}$ & $1402^{+143}_{-110}$ & $1.6\times10^{-3}$ \\
\hline
LRS (GPI+SPHERE) & EddySed (MgSiO$_3$, am) & $0.55^{+0.06}_{-0.14}$ & $0.22^{+0.25}_{-0.47}$ & $0.70^{+0.05}_{-0.03}$ & $5.53^{+0.04}_{-0.05}$ & $1473^{+17}_{-20}$ & 1.0 \\
LRS & EddySed (MgSiO$_3$, cd) & $0.45^{+0.08}_{-0.09}$ & $-0.27^{+0.17}_{-0.19}$ & $0.77\pm0.04$ & $5.45^{+0.04}_{-0.05}$ & $1443\pm28$ & 0.69 \\
LRS & EddySed (MgSiO$_3$ + Fe, am) & $0.66^{+0.07}_{-0.10}$ & $0.21^{+0.18}_{-0.24}$ & $0.73\pm0.03$ & $5.50^{+0.03}_{-0.04}$ & $1458^{+21}_{-19}$ & 1.54 \\
LRS & EddySed (MgSiO$_3$ + Fe, cd) & $0.29^{+0.06}_{-0.07}$ & $-0.51^{+0.17}_{-0.19}$ & $0.75\pm0.03$ & $5.47^{+0.04}_{-0.03}$ & $1453^{+24}_{-21}$ & 2.65 \\
LRS & Clear & $0.12^{+0.02}_{-0.01}$ & $-1.37^{+0.07}_{-0.05}$ & $1.10\pm0.04$ & $5.12\pm0.03$ & $1262\pm16$ & $7.0\times10^{-26}$ \\
\hline
\citet{peretti_orbital_2019} & Cloudy retrieval $^{\rm (a)}$ & $0.13^{+0.14}_{-0.08}$ & $-1.15^{+0.47}_{-0.39}$ & $0.85\pm0.03^{\rm (b)}$ & $5.40\pm0.03$ & $1350\pm50$ & ... \\
\citet{crepp_GPI_2018} & Cloudy grid $^{\rm (a)}$ & ... & ... & ... & $5.2^{+0.5}_{-0.6}$ & $1410^{+130}_{-140}$ & ... 
\enddata
\tablecomments{(a) \citet{peretti_orbital_2019} carried out cloudy retrievals on their SPHERE spectrum (1.0-1.65~$\mu$m) and archival $K$ and $L$ photometry with the HELIOS-R code \citep{lavie_HELIOS_2017}, while \citet{crepp_GPI_2018} fitted their extracted GPI spectrum (1.5-2.2~$\mu$m) to a cloudy grid model \citep{Saumon_2008}. (b) \citet{peretti_orbital_2019} placed a Gaussian prior of $1.0\pm0.1~\Rj$ on the radius. }
\end{deluxetable*}

\begin{figure*}[t!]
    \centering
    \includegraphics[width=0.65\linewidth]{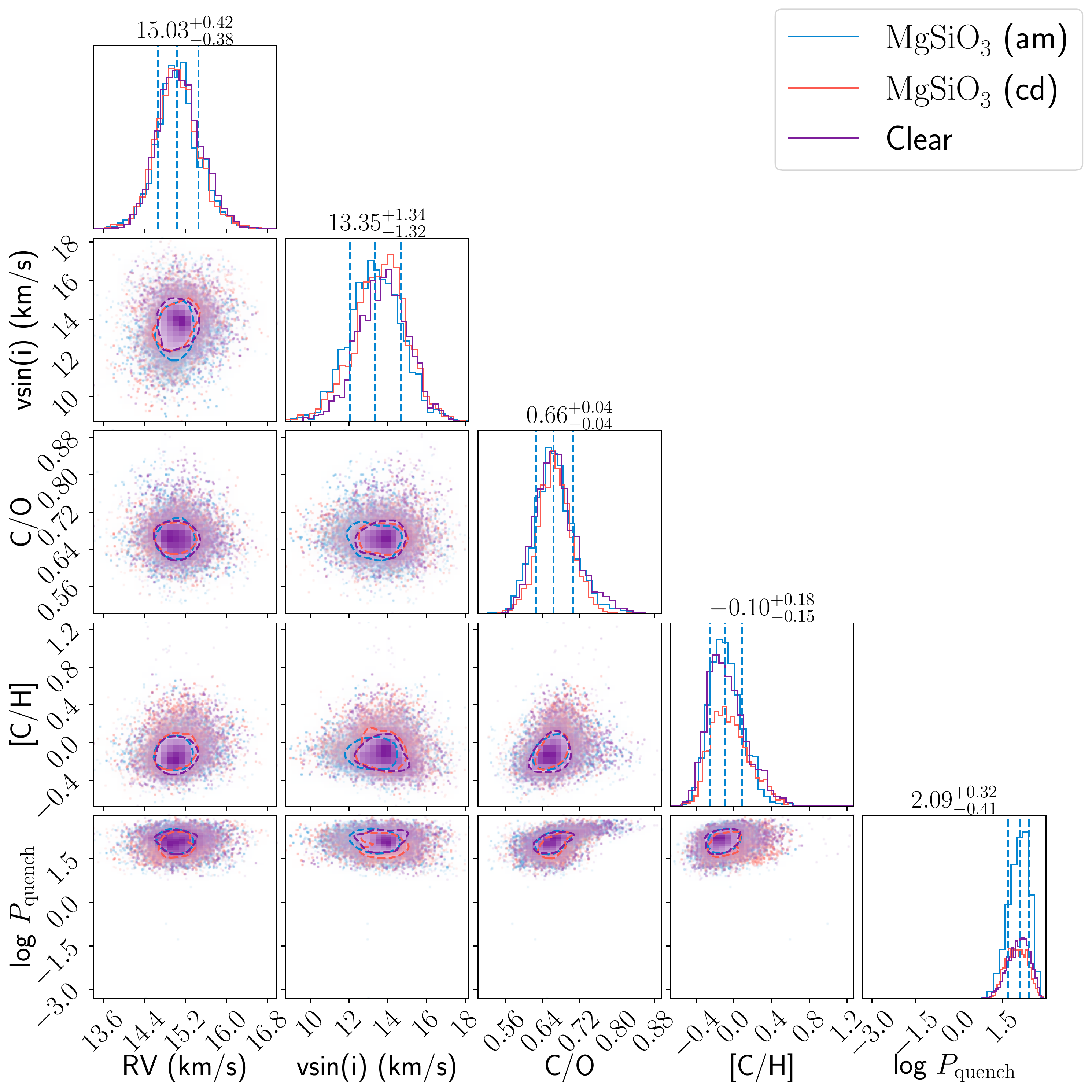}
    \caption{Posterior distributions for five key parameters from HRS retrievals of HD~4747~B, using the EddySed cloud model (MgSiO$_3$, `am' and `cd' in blue and red), and the clear model in purple. The titles on each histogram show the median and 68\% credible interval for the baseline retrieval (MgSiO$_3$, am). Regardless of the cloud model used, the results agree well between different fits for the RV, $v\sin{i}$, C/O, [C/H] (discussed in \S~\ref{sec:overview_abund}), and quench pressure (discussed in \S~\ref{sec:diseq_chem}).}
    \label{fig:kpic_corner}
\end{figure*}

\begin{figure*}[t]
    \centering
    \begin{subfigure}
        \centering
        \includegraphics[width=0.25\linewidth]{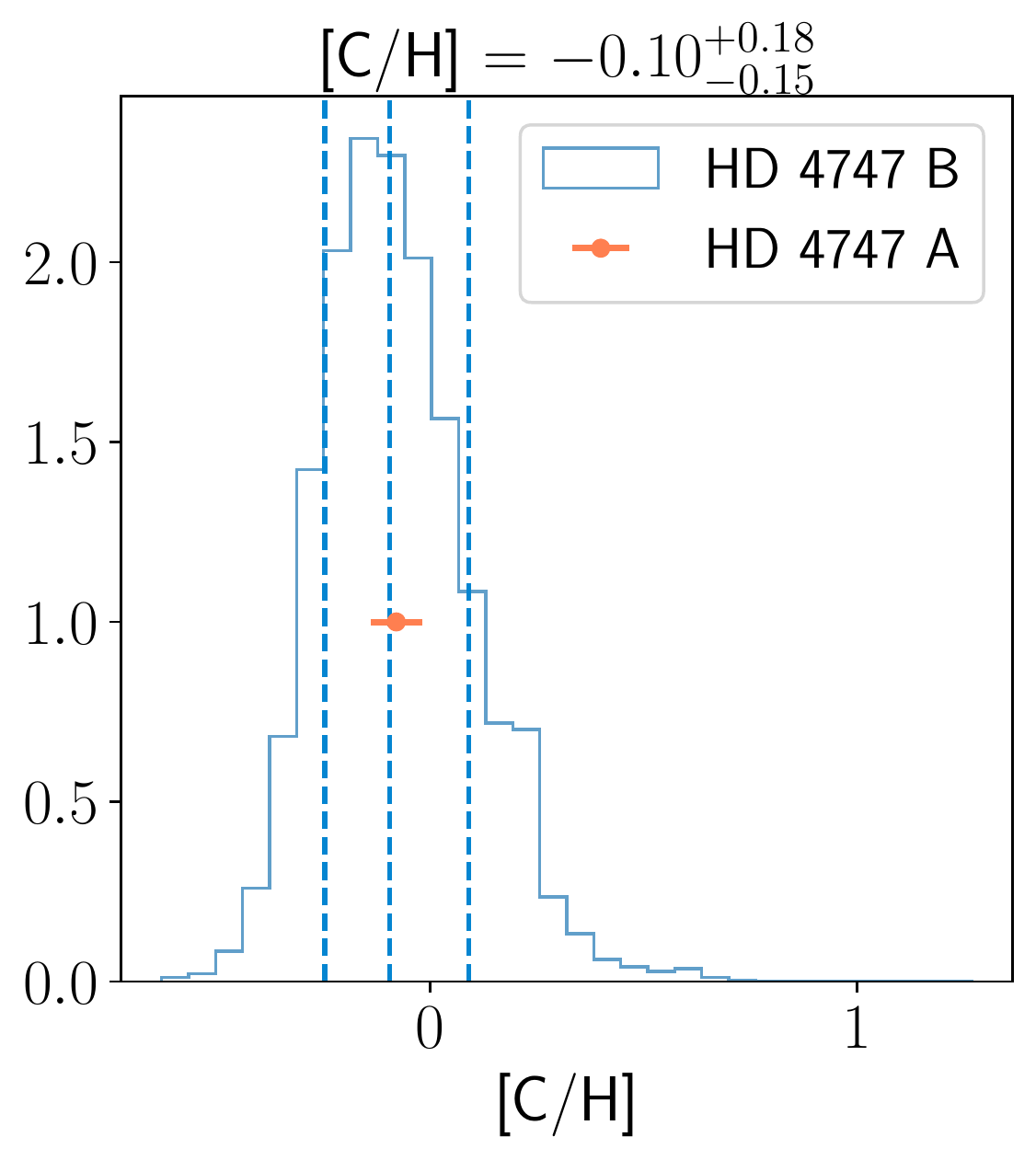}  
    \end{subfigure}
    \begin{subfigure}
            \centering
        \includegraphics[width=0.25\linewidth]{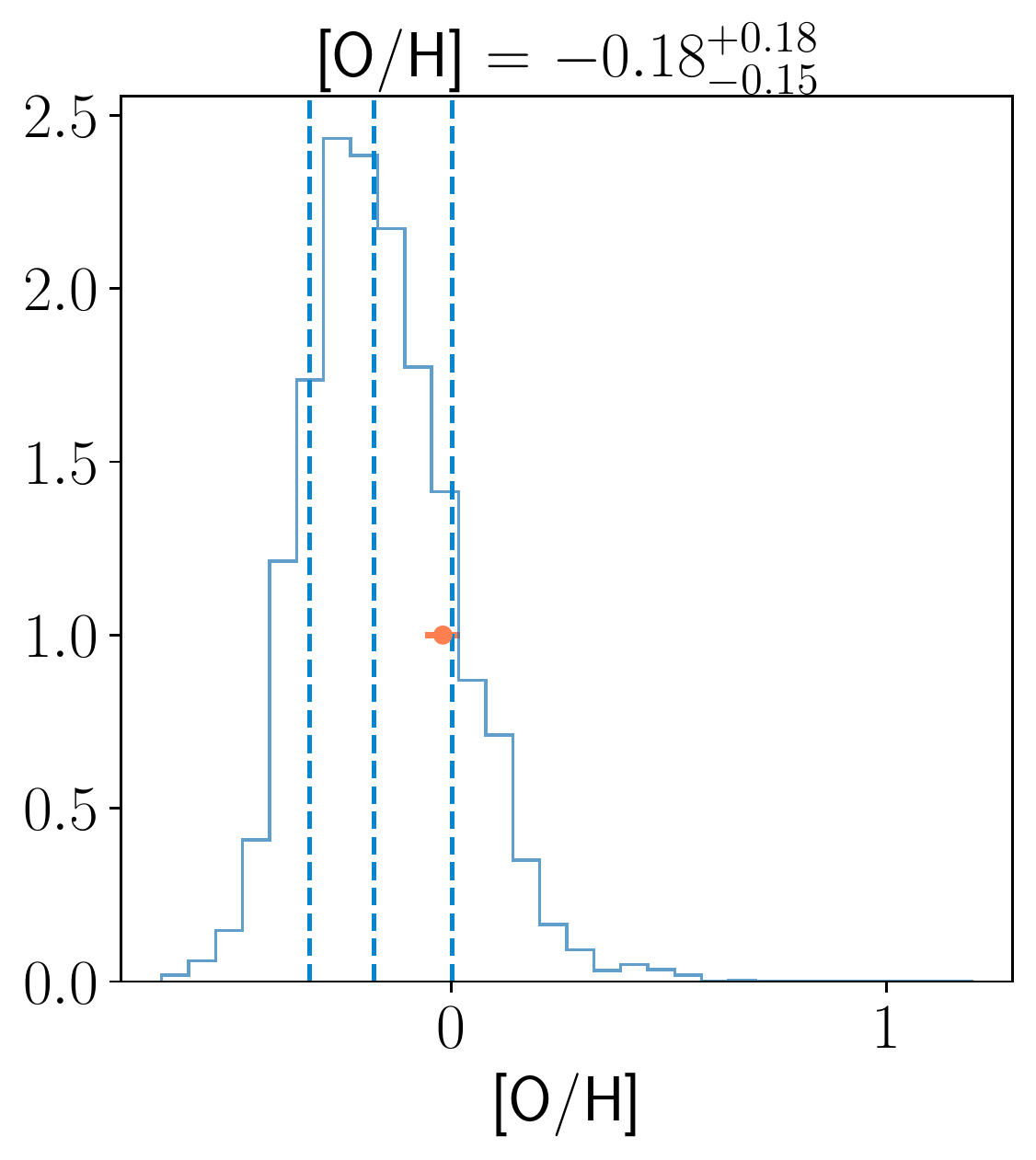}  
    \end{subfigure}
    \begin{subfigure}
            \centering
        \includegraphics[width=0.25\linewidth]{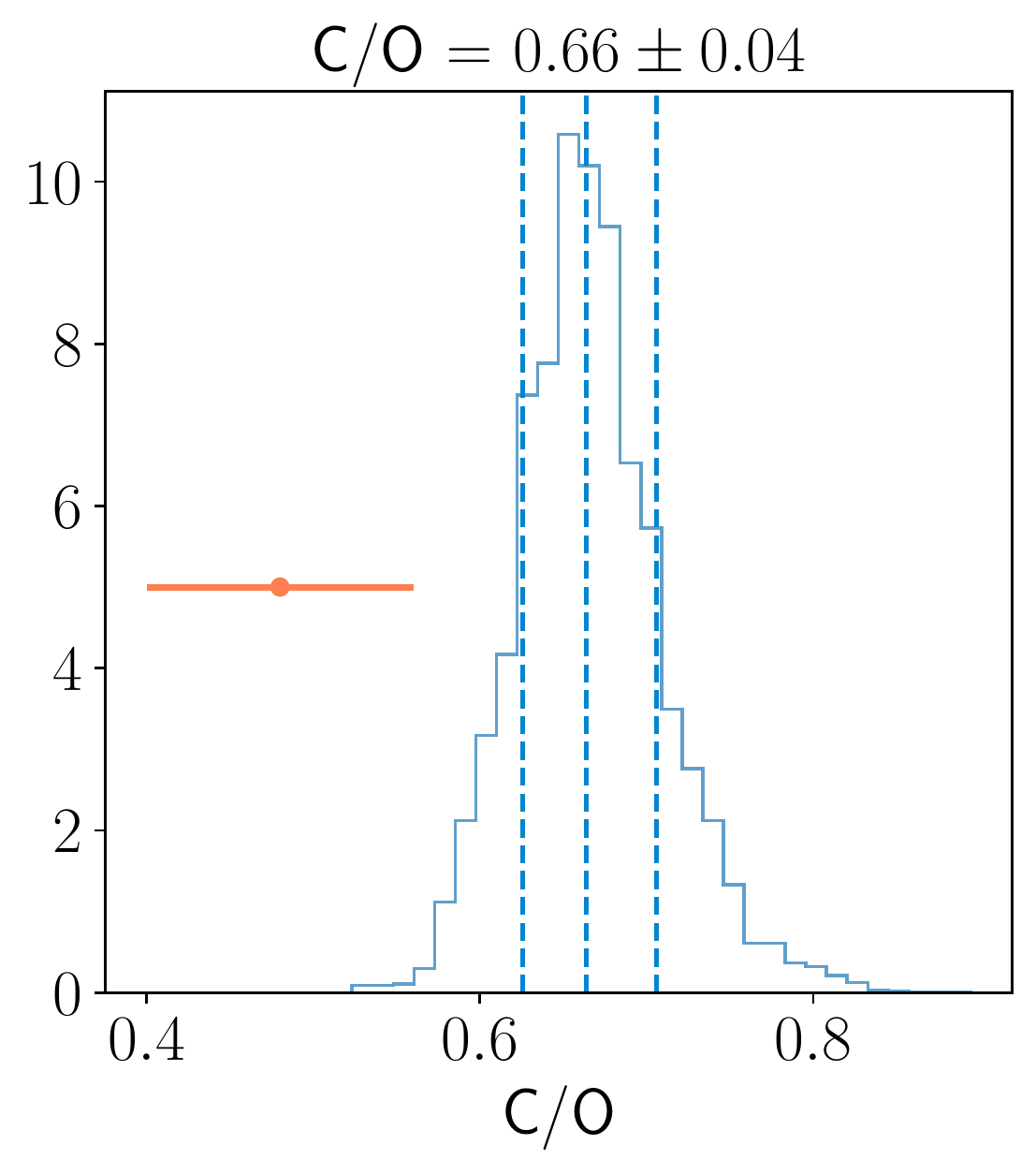}  
    \end{subfigure}
    \caption{Retrieved C and O abundances (relative to solar) and C/O for HD~4747~B in blue. The titles on each histogram showing the median and 68\% credible interval. The red points show the stellar values from \S~\ref{sec:host}. The [C/H] agrees well, [O/H] is consistent within $1\sigma$ between the companion and star, and C/O is consistent at the $2\sigma$ level.}
    \label{fig:c_o_abunds}
\end{figure*}

\section{High-resolution retrievals (KPIC)}\label{sec:hrs_results}

\subsection{Overview} \label{sec:overview_abund}
From our HRS retrievals of HD~4747~B, we find that both clear and cloudy models yield consistent results for the atmospheric parameters (abundances, temperature structure, quenching) and bulk properties (radius, radial velocity, spin). A few selected parameters are plotted in Fig.~\ref{fig:kpic_corner} and tabulated in Table~\ref{table:bayes_factors}. The insensitivity of the HRS retrieval results to clouds, a major finding of this paper, is discussed in \S~\ref{sec:cloud_insensitive}. In Fig.~\ref{fig:kpic_mod_data}, we plot the data, a best fit model, and residuals for the baseline HRS retrieval. We report values from this retrieval as the final results of this paper, with selected parameters shown in the first row of Table~\ref{table:bayes_factors} and joint posterior distributions in Appendix C. We also plot the contribution from the planet and star separately at their best-fit flux levels. We compute the auto-correlation function of the residuals and find that there is no evidence for correlated noise or strong systematics. Unless otherwise specified, we quote results from the baseline EddySed cloud model (MgSiO$_3$, am). See Appendix C for the posterior distributions of other parameters from our baseline model.

To make sure that we are fitting the correct signal, we check the RV and flux level of the companion. From our orbital posteriors for HD~4747~B, the expected RV shift on the night of our HRS observation is $15.0\pm0.1$ km/s in the Earth's reference frame, which is a combination of the system barycenter velocity, the Earth's relative velocity with respect to HD~4747, and the companion's orbital velocity. The fitted RV of $15.0\pm0.4$ km/s agrees perfectly with this value (see Fig.~\ref{fig:kpic_corner}). In addition, the companion flux level in the spectral orders from 2.29 to 2.49~$\mu$m is $85\pm10$ counts, comparable to the speckle flux levels in these orders. Taking into account the difference in wavelengths and the difference in integration time (600~s for the companion, 60~s for the on-axis star), we estimate that our measured companion flux corresponds to $\Delta K_s=8.3\pm0.3$ mag, which is within $3\sigma$ of the photometric $\Delta K_s=9.05\pm0.14$ mag reported by \citet{crepp_GPI_2018}. The agreement between these contrast values are reasonably good given the time-varying throughput of KPIC \citep{delorme_Keck_2021a}, and the fact that we subtract out the continuum with high-pass filtering, effects which complicate a direct flux comparison. 

Fig.~\ref{fig:kpic_corner} also shows the projected spin rate $v\sin{i}=13.2^{+1.4}_{-1.5}$~km/s, which is comparable to the rotation rates observed for field brown dwarfs with similar spectral types \citep[e.g.][]{konopacky_Rotational_2012}. We also plot the retrieved quench pressure $P_{\rm quench}$ in Fig.~\ref{fig:kpic_corner}, which indicates that the chemical reaction timescale becomes longer than the vertical mixing timescale at pressures lower than $P_{\rm quench}$. Thus, disequilibrium chemistry is clearly affecting the atmosphere (see \S~\ref{sec:diseq_chem} for details). 

We compute $T_{\rm eff}$ by sampling from our posterior to generate low-resolution models over a large wavelength range (0.5 to 30~$\mu$m) and calculating the integrated flux. We then solve for $T_{\rm eff}$ using the Stefan-Boltzmann law. When computing $T_{\rm eff}$, we include opacities from Na and K, which are important sources of opacity near visible wavelengths. As shown in Table~\ref{table:bayes_factors}, the retrieved radius and $T_{\rm eff}$ from HRS have broad distributions, which reflect the relatively weak luminosity constraints from the HRS (log($L_{\rm bol}/L_{\odot}$)=$-4.33^{+0.23}_{-0.25}$). This is because the HRS is not flux-calibrated and we remove the continuum in our fits. Comparing to values of radius and $T_{\rm eff}$ from previous work based on LRS \citep{crepp_GPI_2018, peretti_orbital_2019}, our retrieved values from the HRS retrievals are consistent at the $1-2\sigma$ level (see Table~\ref{table:bayes_factors}). We discuss the constraints on these parameters from the LRS in \S~\ref{sec:prior_knowledge}.

We compare our retrieved [C/H], [O/H], and C/O with that of the host star (see \S~\ref{sec:host}) in Fig.~\ref{fig:c_o_abunds}. Our retrieved C abundance agrees well with the host star value, while the O abundance is lower by about $1\sigma$. This results in our retrieved C/O for the companion being higher by about $2\sigma$ compared to the stellar value. Here and elsewhere in the paper, we compute the `$\sigma$ difference' between two measurements by dividing the difference in the two median values by the quadrature sum of the uncertainties from both measurements. We discuss the implications of our measured abundances for HD~4747~B in \S~\ref{sec:discuss_abunds}.

\begin{figure*}[t!]
    \centering
    \begin{subfigure}
      \centering
      \includegraphics[width=0.49\linewidth]{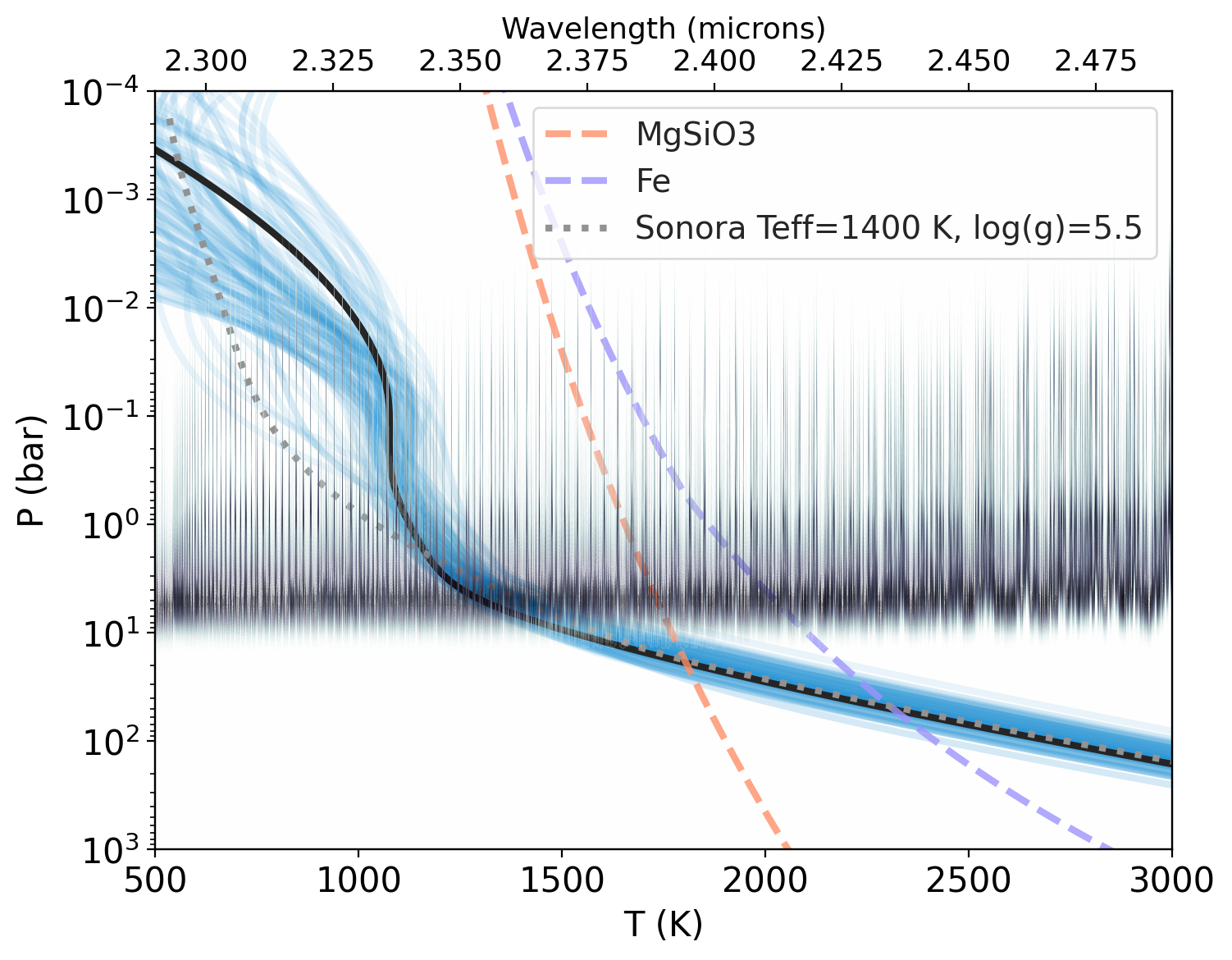}
    \end{subfigure}
    \begin{subfigure}
      \centering
      \includegraphics[width=0.49\linewidth]{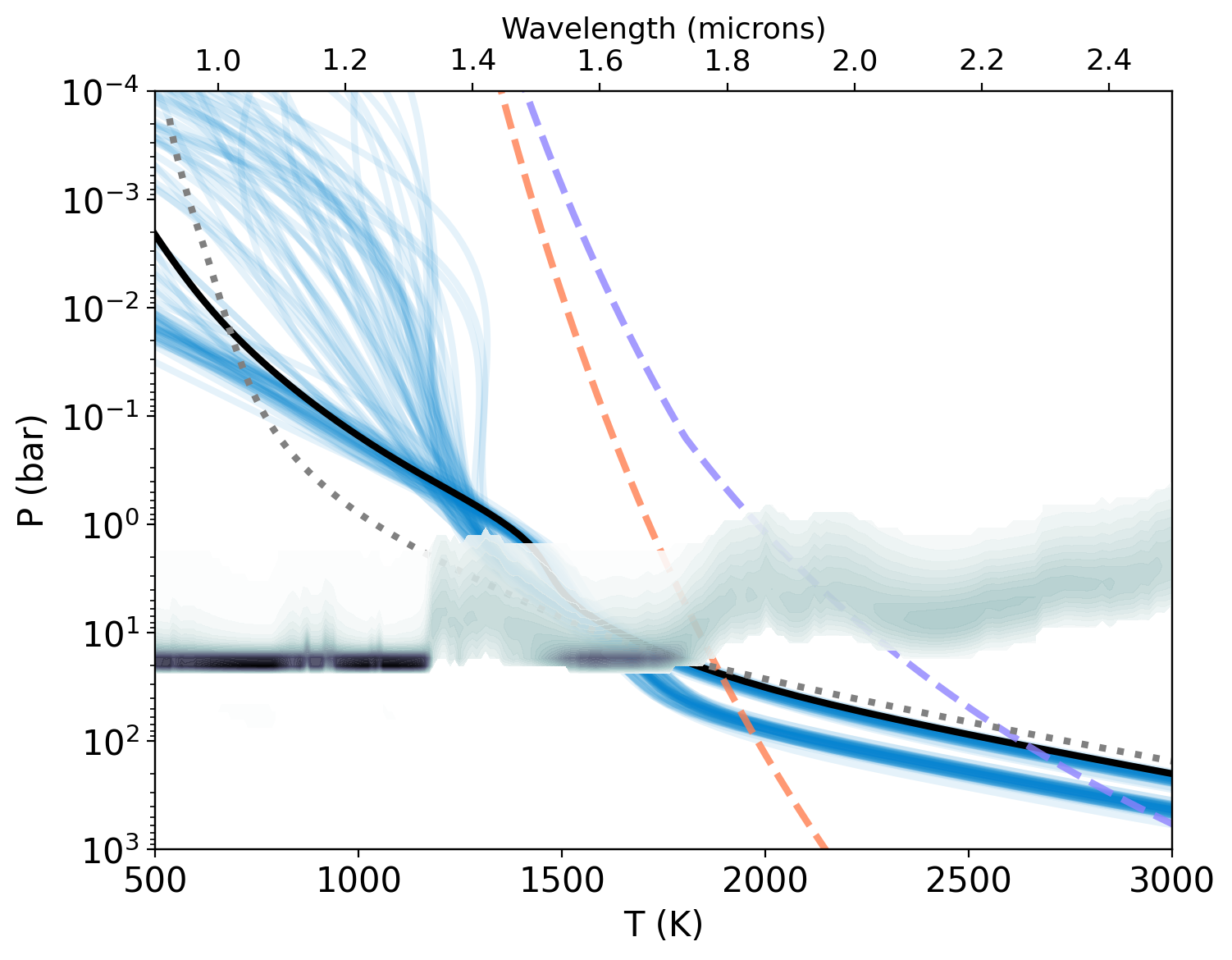}
    \end{subfigure}
    \caption{The P-T profiles from our HRS retrieval (left panel) and LRS retrieval (right panel) with the baseline cloud model (MgSiO$_3$, am). In each panel, black lines show the best-fit profile and blue lines are 100 random draws from the posterior. We also show a cloudless Sonora P-T profile \citep{marley_Sonora_2021} with similar bulk properties as HD~4747~B in dotted gray. The condensation curves for MgSiO$_{3}$ and Fe clouds are plotted in dashed lines. We also overplot the emission contribution function as contours, which show the fraction of flux (darker indicates higher fraction) a given pressure layer contributes to the total flux at a given wavelength \citep{molliere_petitRADTRANS_2019}. Thus, these use the wavelength axes, and not the temperature axes. The HRS is sensitive to the continuum forming around a few bars and line cores which form up to $10^{-2}$ bars. Over the same wavelength range of 2.29-2.49~$\mu$m, the LRS shows the continuum arises from $\approx1-10$~bars, consistent with the HRS.}
    \label{fig:pt_profile}
\end{figure*}

\subsection{Why are our KPIC HRS insensitive to clouds in HD~4747~B?} \label{sec:cloud_insensitive}
Clouds represent a significant source of uncertainty in many published models of substellar atmospheres \citep[e.g.][]{burningham_retrieval_2017, wang_Chemical_2020a}. However, we find that the retrieved parameters from our KPIC HRS are insensitive to the choice of cloud model for HD~4747~B. As shown in Fig~\ref{fig:kpic_corner}, the posteriors for radius, RV, $v\sin{i}$, C/O, [C/H], and quench pressure are nearly identical across the various models. The same is true for other parameters.

Table~\ref{table:bayes_factors} shows that the different cloud models are indistinguishable for the KPIC HRS; the clear model fits as well as the cloudy models, with $B=0.61$, which does not pass the threshold of $B=10/0.1$ to be considered statistically favored/disfavored. This indicates that the data can be fitted adequately without clouds; indeed the cloud parameters for the EddySed models span their respective prior ranges almost uniformly as shown in Appendix C. As we will discuss in \S~\ref{sec:lrs_results}, the LRS show that the atmosphere of HD~4747~B is cloudy. This implies that cloud opacity must be minimal at the pressures probed by our HRS. 

To understand this, we plot in Fig.~\ref{fig:pt_profile} the retrieved P-T profiles (black and blue lines), cloud condensation curves (dashed lines), and emission contribution functions. The left and right panels show results from the HRS and LRS retrievals, respectively. The emission contribution function for HRS shows that we are sensitive to pressures ranging from a few bars, where the continuum forms, up to $\approx10^{-2}$ bars in the cores of individuals lines. Note that the contribution functions use the wavelength axes on the top. In the EddySed model, the cloud base is set at the intersection of the P-T profile and a given cloud condensation curve (dashed lines). For MgSiO$_3$, this corresponds to a pressure of $\approx 10-20$ bars when using our HRS-retrieved P-T profile. As the cloud mass fraction drops exponentially above the cloud base in the EddySed model (controlled by $K_{\rm zz}$ and $f_{\rm sed}$), we find that the cloud opacity decreases to negligible levels by the time we reach pressures of a few bars where the continuum forms. For this reason, we do not consider models with Fe clouds in our HRS retrievals, since the Fe cloud base forms even deeper than that of MgSiO$_3$.

Therefore, our KPIC HRS are insensitive to clouds because we cover both a relatively small wavelength range (2.29-2.49~$\mu$m) and a range where molecular opacities from H$_2$O, CO, and CH$_4$ are significant. The small wavelength range means that the cloud opacity is effectively constant in wavelength. The strong molecular opacity in HRS allows us to resolve many individual absorption lines and obtain good constraints on the atmospheric composition for molecules present in this region of the spectrum. The opacity of these molecules decrease at shorter wavelength due to decreasing excitation cross sections, so the continuum shifts to higher pressures (deeper down) at shorter wavelengths. This effect is visible in the LRS contribution function, where close to $1~\micron$, the emission originates from roughly the same pressure as the MgSiO$_3$ cloud base, making the $y$ and $J$ bands particularly sensitive to clouds (see Fig.~\ref{fig:lrs_fits}).

Could the KPIC HRS be affected by clouds at lower pressures (higher altitudes) than predicted by the EddySed model? Several studies have found that including clouds at lower pressures than predicted by EddySed produces better fits to mid-IR spectra of isolated brown dwarfs \citep[e.g.][]{burningham_Cloud_2021, luna_Empirically_2021}. As shown in Fig.~\ref{fig:pt_profile}, our HRS P-T profiles show a nearly isothermal region between about 0.1-1 bars, which could suggest a degeneracy with clouds \citep{burningham_retrieval_2017}. To check whether the P-T parameterization affects our results, we run a retrieval with a fixed P-T, namely the self-consistent profile over-plotted in gray. We find that all posteriors from this fixed P-T fit overlap within $1\sigma$ with those from our baseline retrieval. Thus, we conclude that the isothermal part of the P-T we retrieve is not biasing our conclusions. To further examine the possibility of clouds at lower pressures, we also run an opaque cloud model with infinite opacity below a retrieved pressure, and a gray cloud model that adds a constant cloud opacity at each pressure layer. When fitting the HRS with these more flexible cloud models, we also find consistent results with the baseline model. In the second model, the gray opacity is bounded to lie below $\sim0.03~g/\rm{cm}^{-3}$, and the pressure of the infinitely opaque cloud is required to be deeper than $\sim1$ bar. Therefore, even with these less constraining cloud parameterizations, we find that our HRS still prefers solutions with minimal cloud opacity.

\subsection{Disequilibrium chemistry with deep quenching pressure} \label{sec:diseq_chem}
In our retrievals, we include a simple model for disequilibrium chemistry using the quench pressure prescription in \texttt{petitRADTRANS}, which is motivated by \citet{Zahnle_methane_2014}. Specifically, the abundances of CH$_4$, CO, and H$_2$O are held constant at atmospheric pressures lower than the retrieved $P_{\rm quench}$ parameter. We find that when including quenching, the goodness of fit increases drastically compared to fits with full chemical equilibrium. For example, between two clear retrievals with and without quenching, we find that $B\approx380$ in favor of the quenched retrieval. From the Bayes factor interpretation of \citet{benneke_How_2013}, this represents a detection of quenching at $\approx3.9\sigma$ significance. The quench pressure retrieved is also highly consistent between retrievals with different cloud models, with 1 and 2$\sigma$ intervals of $50-260$ and $14-836$ bars (Fig.~\ref{fig:kpic_corner}). In this section, we explore reasons why the data prefer disequilibrium chemistry in the atmosphere of HD~4747~B. The physical implications of our retrieved quench pressure, including an estimate of the vertical diffusion coefficient ($K_{\rm zz}$), are discussed in \S~\ref{sec:discuss_diseq_chem}.

\begin{figure}[t]
    \centering
    \includegraphics[width=\linewidth]{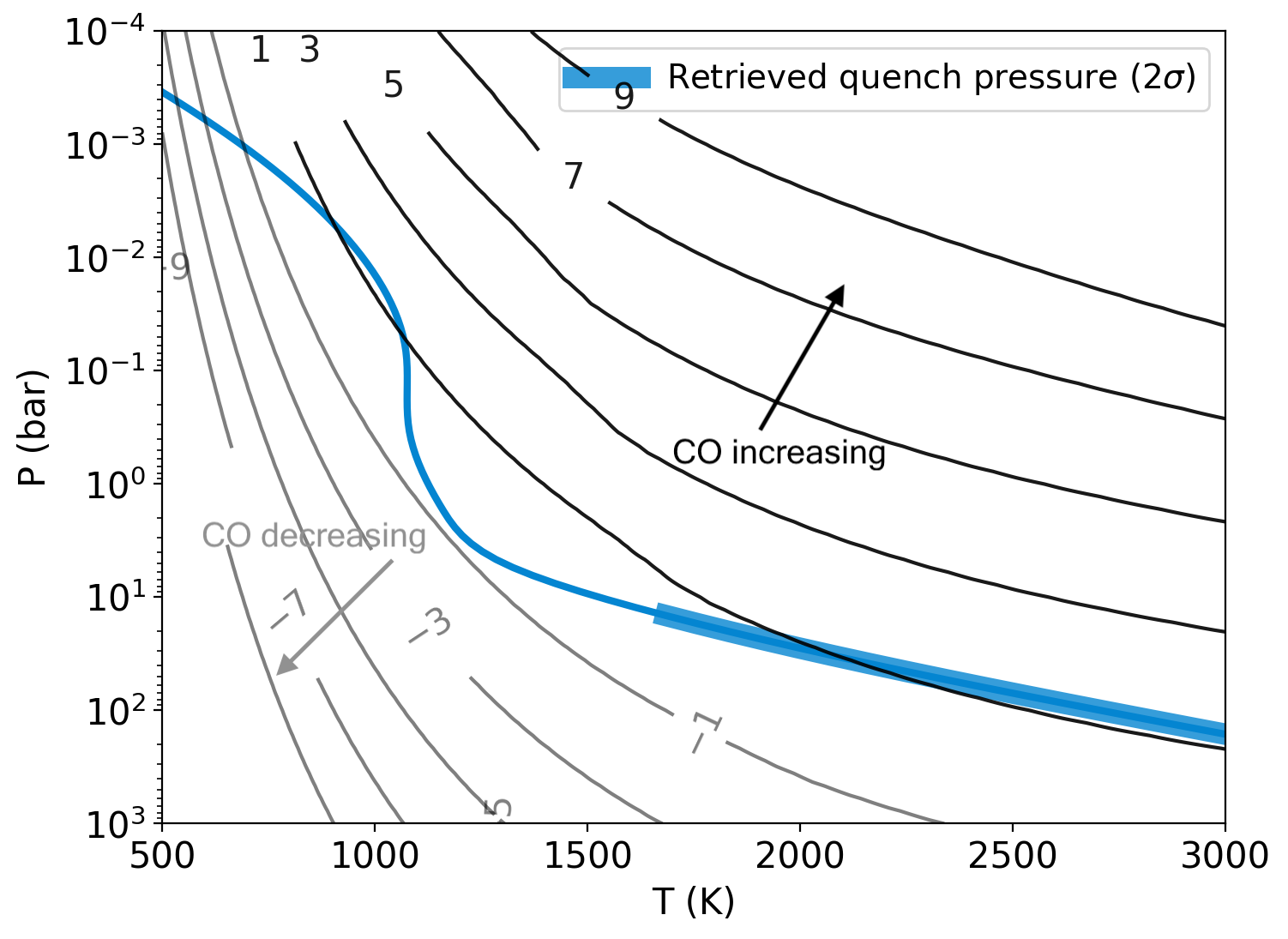}
    \caption{The best-fit P-T profile from our baseline HRS retrieval in blue, with the thicker region indicating the retrieved quench pressure ($2\sigma$ interval of 14-836 bars). Lines of constant log(CO/CH$_4$) volume mixing ratios are shown, with black lines (gray lines) indicating the region where CO (CH$_4$) is more abundant. The P-T profile nearly overlaps with the log(CO/CH$_4$) = +1 line below $\sim20$ bars, which is where we retrieve the quench pressure to be from the HRS.}
    \label{fig:pt_profile_coch4}
\end{figure}

\begin{figure}[t]
    \centering
    \includegraphics[width=\linewidth]{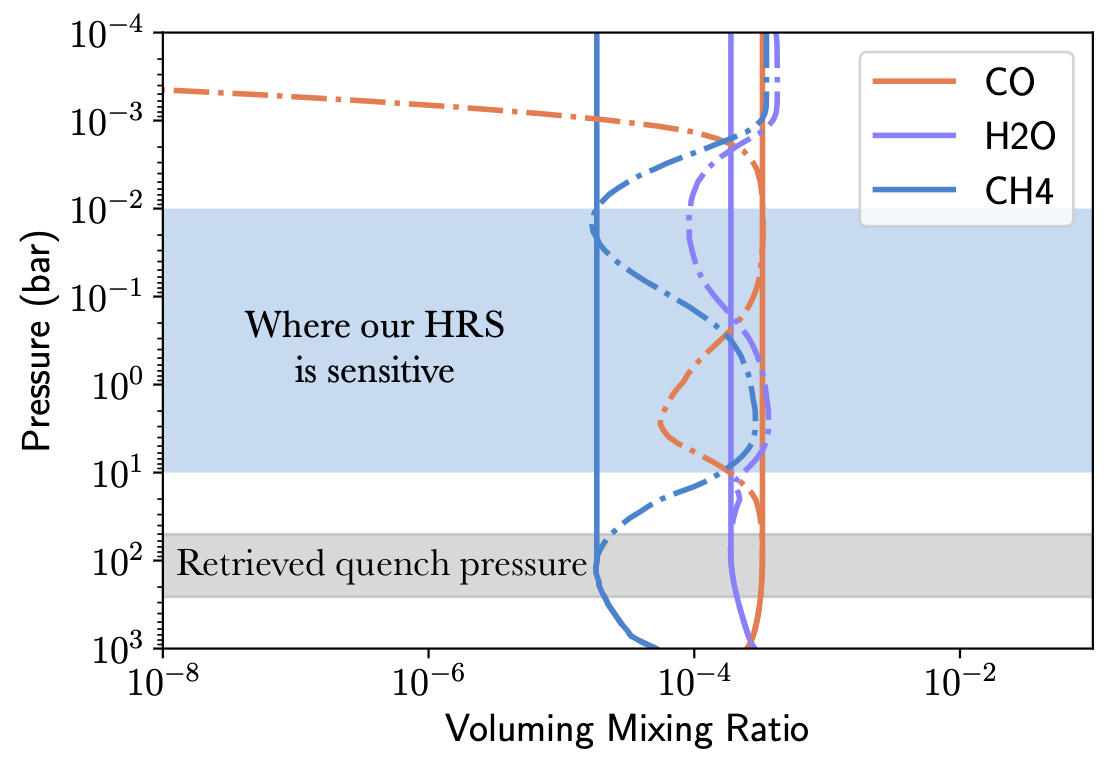}
    \caption{Solid lines: best fit volume mixing ratios (VMRs) of CO, H$_2$O, and CH$_4$ from our baseline HRS retrieval with chemical quenching. Dashed lines: the corresponding VMRs when quenching is turned off (i.e. equilibrium abundances) for the same retrieval. The $1\sigma$ quench pressure is indicated by the gray region, and the blue region shows schematically the pressures where our HRS is sensitive. Within the blue region, the relative CO/CH$_4$ ratio can differ by orders of magnitude between the quenched abundances and the equilibrium abundances.}
    \label{fig:spaghetti_plot_kpic}
\end{figure}

To understand why the data prefer a deep quench pressure, we plot lines of constant log(CO/CH$_4$) volume mixing ratios (VMR) along with the best-fit P-T profile from our baseline HRS retrieval in Fig.~\ref{fig:pt_profile_coch4}. We calculate CO/CH$_4$ from this quenched chemistry retrieval by finding the abundances of each molecule in the chemical grid, iterating over our posterior distribution of C/O, [C/H], and P-T profile. We find that CO/CH$_4$=$13.6^{+5.8}_{-4.6}$. If the atmosphere was in chemical equilibrium, we repeat our calculation and find that we would expect CO/CH$_4$=$1.35^{+0.21}_{-0.17}$, which is ten times smaller than our retrieved value in the quenched chemistry model. Thus, the relative under-abundance of CH$_4$ relative to CO in our HRS leads our models to prefer a deep quench pressure. The value of CO/CH$_4$ also determines our retrieved the quench pressure, whose $2\sigma$ interval is indicated by the thick blue region in Fig.~\ref{fig:pt_profile_coch4}. Because the P-T profile nearly overlaps the curve of CO/CH$_4$ = 10 at $\sim 20$ bars and deeper, a broad range of quench pressures deeper than $\sim20$ bars are consistent with the data. 

\begin{figure*}[t!]
    \centering
    \begin{subfigure}
      \centering
      \includegraphics[width=0.35\linewidth]{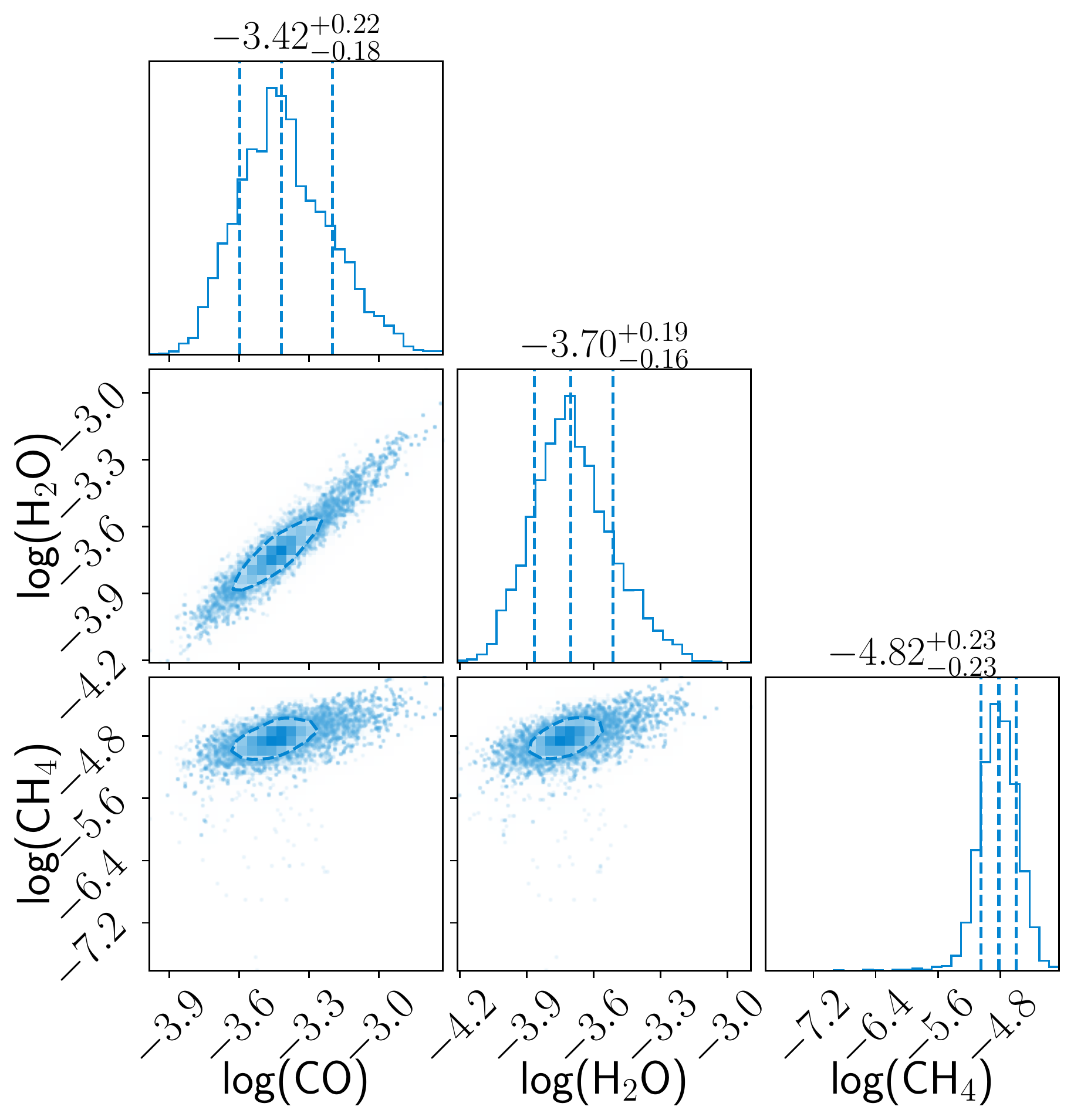}
    \end{subfigure}
    \begin{subfigure}
      \centering
      \includegraphics[width=0.4\linewidth]{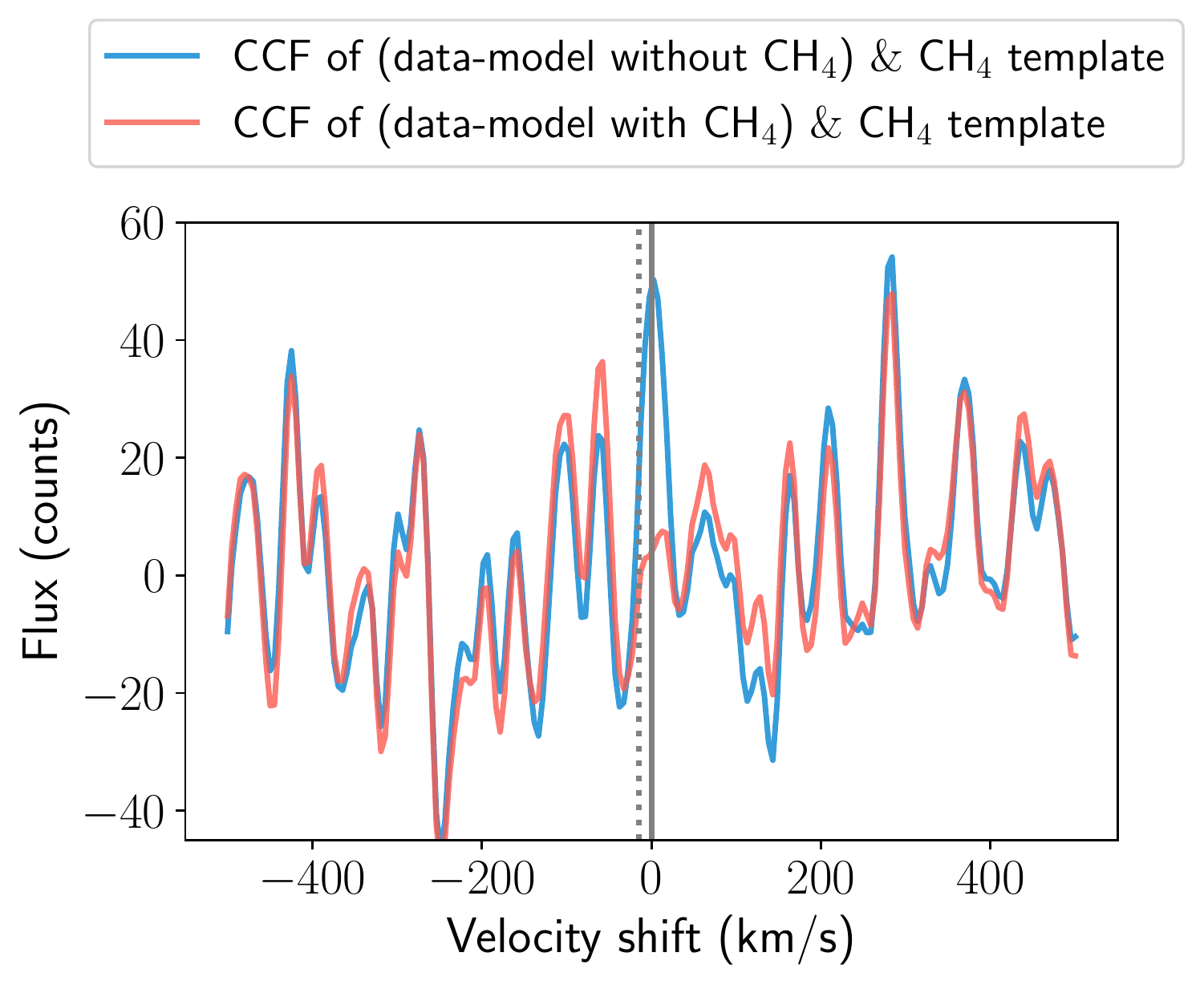}
    \end{subfigure}
    \caption{Left panel: Joint posterior distributions of the log(VMR) of CO, H$_2$O, and CH$_4$ from a KPIC HRS retrieval where we directly fit the molecular abundances and assume they are constant across pressure. Right panel: Cross-correlation functions of a pure CH$_4$ template with (KPIC data - model without CH$_4$) in blue, and the CCF of the CH$_4$ template with (data - model with CH$_4$) in red. The CH$_4$ template is generated with best fit parameters of the full model with CH$_4$, CO, and H$_2$O and manually setting opacities of CO and H$_2$O to zero. The gray solid line indicates the companion rest frame and the gray dotted line is the telluric rest frame. The residuals are taken from two spectral orders (2.29-2.41~$\mu$m) with stronger CH$_4$ detection.}
    \label{fig:ch4_detect}
\end{figure*}

As another way to visualize the detection of disequilibrium chemistry, we plot the molecular abundances in VMR as a function of pressure in Fig.~\ref{fig:spaghetti_plot_kpic}. The solid lines show the VMRs for the HRS quenched chemistry retrieval, while dashed lines show the VMRs for the same model with quenching turned off manually. By comparing the solid and dashed lines, we see that at the pressures probed by our observations, the relative abundances of CO, CH$_4$, and H$_2$O differ by several orders of magnitude between the quenched model and expectations from chemical equilibrium. 

\subsection{Detection of methane in the HRS} \label{sec:ch4_detect}

\begin{deluxetable}{ccccc}
\tablecaption{Free retrievals carried out on HD~4747~B for validating the CH$_4$ detection in the KPIC HRS (Sec~\ref{sec:ch4_detect}). We list the log volume-mixing ratios of molecules included, and the Bayes factor between the model with and without CH$4$.} \label{tab:free_retr}
\tabletypesize{\footnotesize}
\tablehead{\colhead{Molecules} & \colhead{log(CO)} & \colhead{log(H$_2$O)} & \colhead{log(CH$_4$)} & \colhead{$B$} } 
\startdata
CO, H$_2$O & $-3.51^{+0.21}_{-0.17}$ & $-3.77^{+0.19}_{-0.16}$ & ... & 1 \\
CO, H$_2$O, CH$_4$ & $-3.42^{+0.22}_{-0.18}$ & $-3.70^{+0.19}_{-0.16}$ & $-4.82\pm0.23$ & 84
\enddata
\end{deluxetable}

In this section, we take a closer look at the relatively weak methane absorption signal in our HRS, which leads us to prefer quenched models where the CO/CH4 ratio is a factor of ten higher than predicted in models assuming chemical equilibrium. We confirm the presence of detectable levels of methane in the HRS by running a pair of free retrievals, one with only H$_2$O and CO, and one with H$_2$O, CO, and CH$_4$. The results of these retrievals are listed in Table~\ref{tab:free_retr}. In these free retrievals, we fit the abundances of each absorbing species independently and assume a constant abundance as a function of pressure. Although we also considered models that included NH$_3$ and CO$_2$, we only obtained upper limits on their abundances, and therefore excluded them from our fits in this section. Finally, given the insensitivity of the HRS to clouds, we carry out these tests with the clear model to save computation time. 

We find that the data strongly prefer the model with CH$_4$, with a Bayes factor of 84 ($3.4\sigma$ significance; \citealt{benneke_How_2013}). As shown in Table~\ref{tab:free_retr}, we obtain log(CH$_4$)$=-4.82\pm0.23$ from the free retrieval, and the CH$_4$ posterior in Fig.~\ref{fig:ch4_detect} shows no strong covariance with the abundances of either CO or H$_2$O. If the atmosphere was in chemical equilibrium, we would expect a CH$_4$ VMR that is ten times higher than what we retrieve, according to the same calculation described in \S~\ref{sec:diseq_chem}. We note that the abundances from the free retrieval with CH$_4$ also agree well with the corresponding VMRs from our quenched chemistry retrievals. This is not surprising given the deep quench pressure we retrieve, which makes the molecules abundances constant in the regions where our HRS is sensitive (see Fig.~\ref{fig:spaghetti_plot_kpic}).

We separately visualize the CH$_4$ detection in cross-correlation space by carrying out an analysis similar to that described in \citet{zhang_12CO_2021}. First, we make a `pure CH$_4$ template' from the best-fit companion model with CH$_4$, H$_2$O, and CO by manually setting the abundances of H$_2$O and CO to zero. If the model without CH$_4$ is fitting poorly due to its inability to fit CH$_4$ lines in the data, we would expect the residuals of this model, which we denote $R$ = (data - model without CH$_4$), to contain CH$_4$ lines. Therefore, we cross-correlate $R$ with the pure CH$_4$ template, plotted as the blue CCF in Fig.~\ref{fig:ch4_detect}. In addition, we plot the CCF of $R^\prime=$ (data - model with CH$_4$) with the pure CH$_4$ template in red for comparison. The blue CCF shows a peak at 0 km/s (solid gray line), where we expect a real signal to be since the models have been shifted by the best fit companion RV. If the residuals were dominated by telluric CH$_4$ for example, the CCF peak would appear at the dotted gray line (negative of the RV, or -15 km/s). Thus, even though the height of the CH$_4$ peak in the blue CCF is small compared to the surrounding structure, the fact that it is located at the companion RV is evidence of a real signal from CH$_4$. 

In our CCF framework, the y-axis is the estimated flux level (in counts) of the companion signal from a least-squares minimization. As shown in Fig.~\ref{fig:ch4_detect}, we find a flux level of $\approx50$ counts for CH$_4$, which is an estimate of the companion flux in the residuals. Importantly, this value is consistent with the flux value found when we repeat the same CCF analysis with H$_2$O (i.e. comparing a model with only CO and CH$_4$ and the baseline model of CO, H$_2$O, and CH$_4$). For a molecule such as NH$_3$, which we see no evidence of in the KPIC HRS, the flux value from the CCF becomes unbounded as the least-squares routine used for computing the CCF fails to converge. 

Finally, we check for cross-talk between H$_2$O and CH$_4$ by cross-correlating $R$ with the pure water template and detect no CCF peaks. Furthermore, we note that in a retrieval with only CO and CH$_4$ (no H$_2$O), the retrieved CH$_4$ abundance is consistent with the value from the full model including CO, H$_2$O and CH$_4$. 

We therefore conclude that the data strongly favor the presence of detectable levels of methane in the HRS, with an abundance significantly lower than that predicted by equilibrium chemistry models. The detection of methane at log(CH$_4$) = ${-4.82\pm0.23}$ demonstrates the ability of KPIC to retrieve species that are more than an order of magnitude lower in VMR than the dominant molecular constituents in the data, in only 1 hour of integration time.

\begin{figure*}
\centering
  \includegraphics[width=0.75\linewidth]{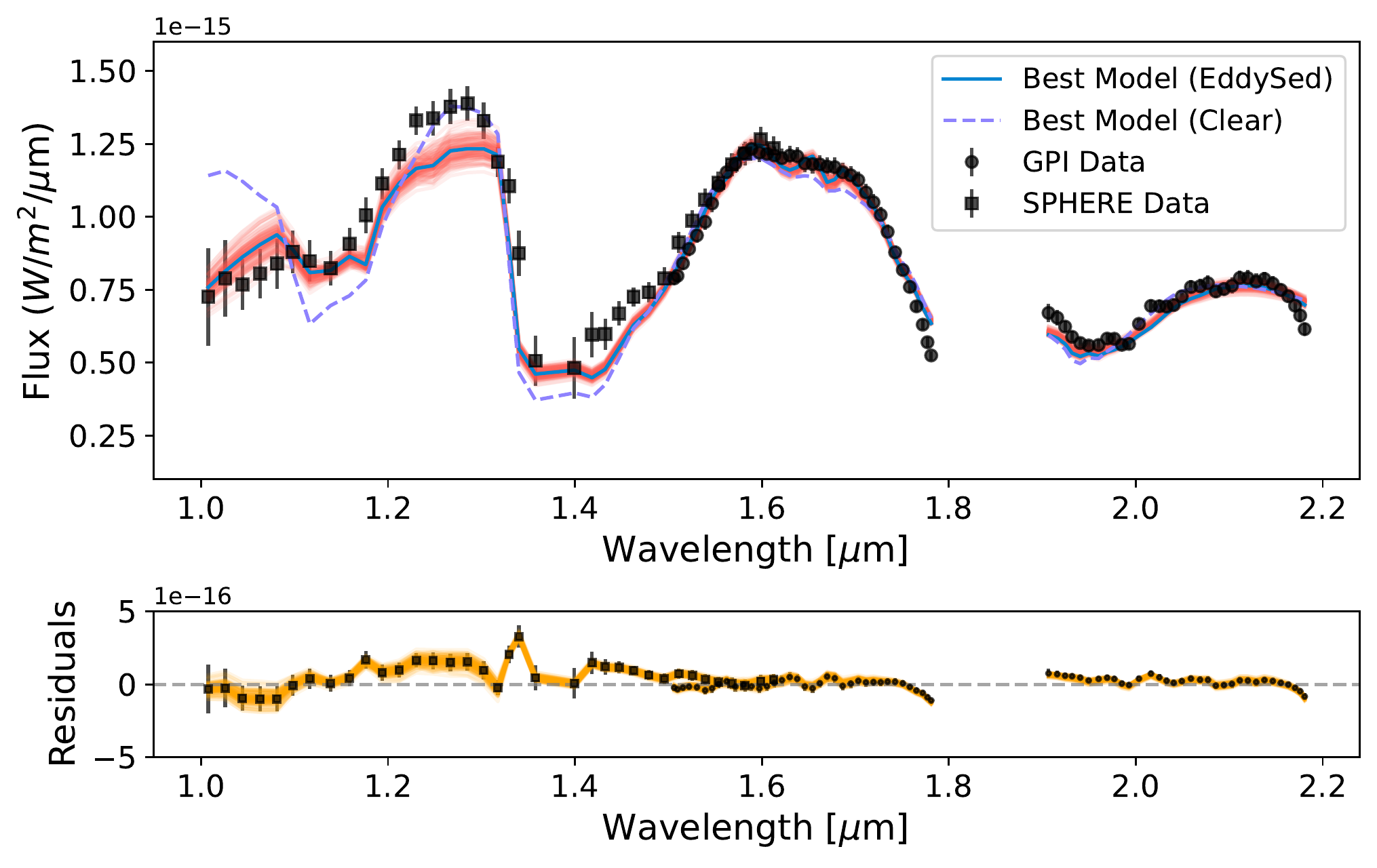}
\caption{Top panel: best-fit cloudy model (EddySed + MgSiO$_3$, am) in blue and random models drawn from the posterior in light orange for a LRS retrieval of HD~4747~B. The best-fit clear model is overplotted in dashed purple, which fits visibly worse from $\approx1.0-1.2~\mu$m.  Bottom panel: the residuals of the cloudy model shown in error bars, along with random draws of the GP models for the residuals in orange.}
\label{fig:lrs_fits}
\end{figure*}

\section{Low-resolution retrievals (GPI + SPHERE)}\label{sec:lrs_results}

\subsection{Overview}
In this section, we present the results from our fits to the LRS and compare our retrieved parameters to those from the HRS fits. We fit the LRS using the same models as before. These include one clear model and four different implementations of the EddySed cloud model where we vary our assumptions about the unknown cloud properties. The cloudy models consist of two MgSiO$_3$ retrievals with am and cd particles (explained in \S~\ref{sec:cloud_choice}), and two retrievals with MgSiO$_3$ and Fe clouds (again, am and cd). In Fig.~\ref{fig:lrs_fits}, we plot the data, best-fit cloudy and clear models, the residuals, and the GP models of the residuals. The posteriors for a few key parameters from these retrievals are plotted in Fig.~\ref{fig:lowres_corner} and tabulated in Table~\ref{table:bayes_factors}. See Appendix C for the posterior distributions of other parameters in the baseline model.

When comparing the clear and cloudy models in Fig.~\ref{fig:lrs_fits}, we see that the data shortward of $\sim1.2~\mu$m is poorly fit by models without clouds. This causes the clear model to have $B\approx7\times10^{-26}$; it is overwhelmingly ruled out compared to the baseline EddySed model. In addition, when we plot the models over a larger wavelength range in Fig.~\ref{fig:lowres_long_wave}, we find that the cloudy models agree with the NIRC2 L band photometry from \citet{crepp_TRENDS_2016}, while the clear model over-predicts the $L$ band flux by $\approx2\sigma$. We did not include these photometric points in our retrievals. 

Fig.~\ref{fig:lrs_fits} shows that the SPHERE J band data from $\approx1.2-1.35~\mu$m is not well fit by even the cloudy model, which could either be caused by model mismatch or speckle contamination that artificially raises the flux. The GP model finds that $\sim60\%$ of the SPHERE error bars and $\sim90\%$ of the GPI error bars are from correlated noise, with correlation length scales of $\sim6$ and $\sim2$ wavelength channels, respectively. This confirms our initial intuition that the noise in the SPHERE and GPI images is likely dominated by correlated speckle noise based on visual inspection of the images. For the SPHERE data set, we estimate that the retrieved length scale is roughly equal to the number of steps that a speckle would move across the PSF for our brown dwarf's separation; indeed, we see speckles moving across the companion PSF in the reduced images. Overall, the SPHERE spectrum is less reliable than that from GPI because only 4 exposures are available, compared to the $\sim40$ exposures from GPI.

Finally, the P-T profile retrieved from our baseline LRS retrieval shows a bi-modal distribution (see Fig.~\ref{fig:pt_profile}). The degeneracy seen here may be related to issues with the LRS (see \S~\ref{sec:shortcoming_lrs}).

\begin{figure}[t!]
    \centering
    \includegraphics[width=\linewidth]{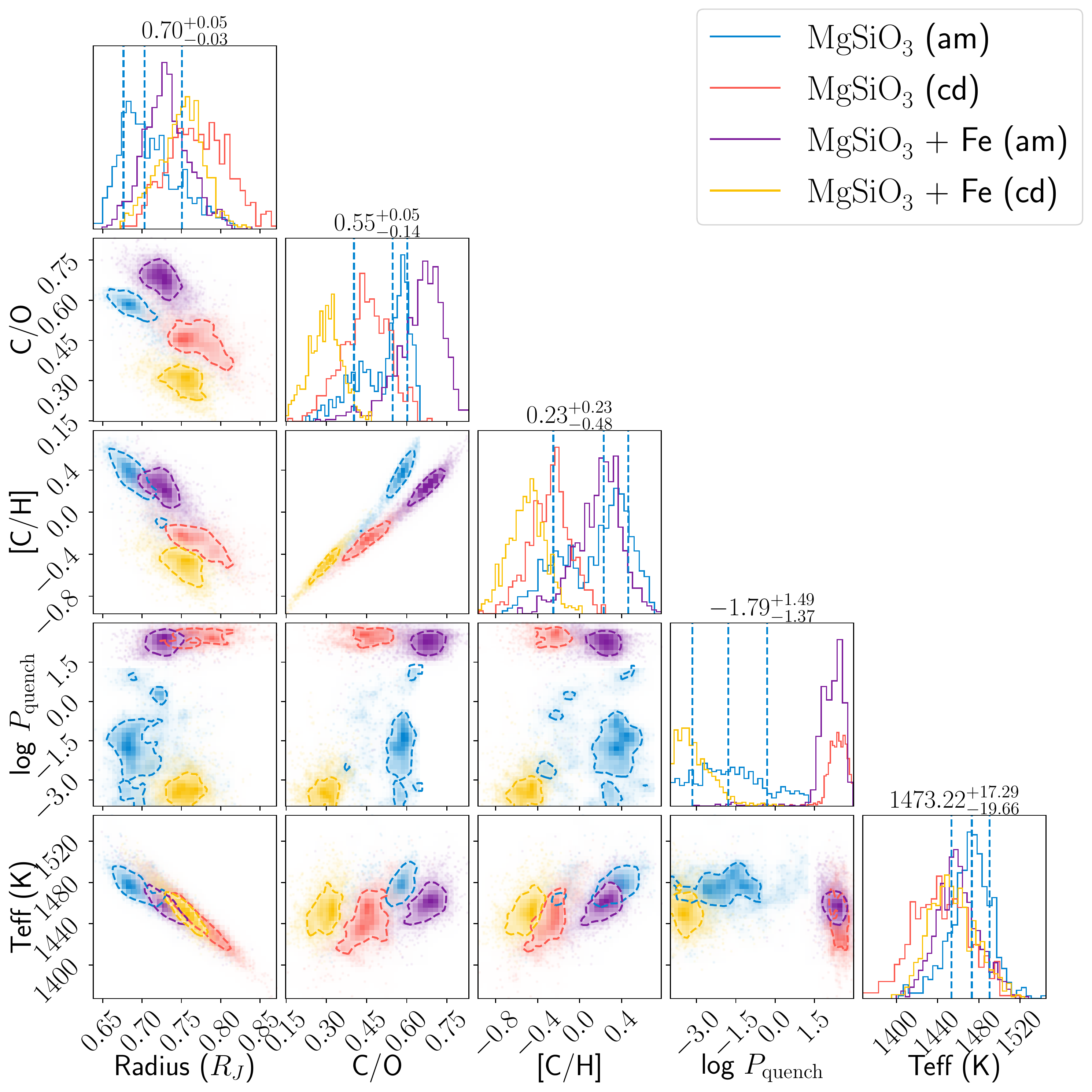}
    \caption{Posterior distributions for a few key parameters from LRS retrievals of HD~4747~B, using the EddySed model with MgSiO$_3$ clouds (blue: amorphous; red: crystalline), and MgSiO$_3$ + Fe clouds (purple: amorphous, yellow: crystalline). The titles on each histogram show the median and 68\% credible interval for the MgSiO$_3$, am model. The results disagree by as much as $3\sigma$, especially in 2-D space, and display strong covariance between C/O and [C/H]. The radius retrieved is also generally smaller than predicted by evolutionary models.}
    \label{fig:lowres_corner}
\end{figure}

\subsection{Comparison with prior knowledge}\label{sec:prior_knowledge}
Because the LRS is flux-calibrated, we can check whether our retrieved radii and effective temperatures are physical and consistent with prior knowledge for this benchmark companion. Using the known age and mass of HD~4747~B ($3\pm2$ Gyr and $m=67.2\pm1.8$~\Mj), we interpolate the COND evolutionary models \citep{baraffe_evolutionary_2003} to find a model-predicted radius of $0.8^{+0.07}_{-0.03}\Rj$, and a predicted $T_{\rm eff} = 1450^{+350}_{-180}$~K. \citet{peretti_orbital_2019} compared the SPHERE spectrum of HD 4747B to those of field brown dwarfs to derive a more tightly constrained $T_{\rm eff}=1350\pm50$~K (see Table~\ref{table:bayes_factors}), which we adopt in the subsequent discussion.

We calculate the effective temperatures of the models in our retrievals by integrating the flux over 0.5-30~$\mu$m. For the baseline EddySed model, we find $T_{\rm eff}=1473^{+17}_{-20}$~K, and a radius of $0.70^{+0.05}_{-0.03}~\Rj$. Compared to prior expectations however, the radius retrieved is too small by $\approx1.5\sigma$ while $T_{\rm eff}$ is too high by $\approx2\sigma$. From substellar evolutionary models, the minimum possible radius of a brown dwarf should be $\approx0.74\Rj$, which is imposed by electron degeneracy pressure \citep{chabrier_massradius_2009}. We find that $T_{\rm eff}$ and radius are correlated in the LRS retrievals, as shown in Fig.~\ref{fig:lowres_corner}, which is expected as different combinations of these two parameters can produce the same total luminosity. However, our total luminosity agrees well with the luminosity predicted by evolutionary models.

Several previous retrieval studies have also found smaller than expected radius for L dwarfs, which may be attributed to the presence of heterogeneous surface features, such as patchy clouds, that are not captured in current 1-D retrieval frameworks \citep[e.g.][]{kitzmann_Heliosr2_2020, gonzales_Retrieval_2020, burningham_Cloud_2021}. On the other hand, \citet{gonzales_First_2021} retrieved a radius consistent with evolutionary models for a seemingly cloudless L dwarf. Whether the radii from evolutionary models are correct is an assumption that is now being tested by a growing sample of transiting brown dwarfs from TESS \citep[e.g.][]{carmichael_Two_2020}. 

In our retrievals with both MgSiO$_3$ and Fe clouds, we retrieve slightly larger radii that are more consistent with the evolutionary model prediction. This could indicate that a single cloud model (MgSiO$_3$) may be inadequate in attenuating the flux from the deep atmosphere. However, models with two cloud species do not improve the fit significantly ($B$ = 1.5-3 compared to the baseline model with MgSiO$_3$ only). Furthermore, the MgSiO3, cd model actually has the largest retrieved radius, but our data cannot distinguish between crystalline and amorphous particles. We conclude that our retrieved radius is sensitive to aspects of the cloud models that are poorly constrained by the existing data for this object.

\subsection{LRS at longer wavelengths could improve abundance and cloud constraints} \label{sec:shortcoming_lrs}

While the LRS can provide tighter constraints on the cloud parameters and radius compared to the HRS, we find that many retrieved parameters, including the atmospheric abundances, are very sensitive to model choices. In Fig.~\ref{fig:lowres_corner}, we overplot the posteriors distributions of a few parameters from our four EddySed models. The retrieved C/O and [C/H] have large uncertainties and can disagree at the $3\sigma$ level between models. The values also span a significant portion of the parameter space ($>$1 dex in metallicity), and show much stronger covariance compared those measured from the HRS (see Fig.~\ref{fig:kpic_corner}). However, all cloudy models fit the LRS well, with Bayes factors within a factor of $\sim3$ (see Table~\ref{table:bayes_factors}), so we cannot distinguish between them.

\begin{figure}[t!]
    \centering
    \includegraphics[width=\linewidth]{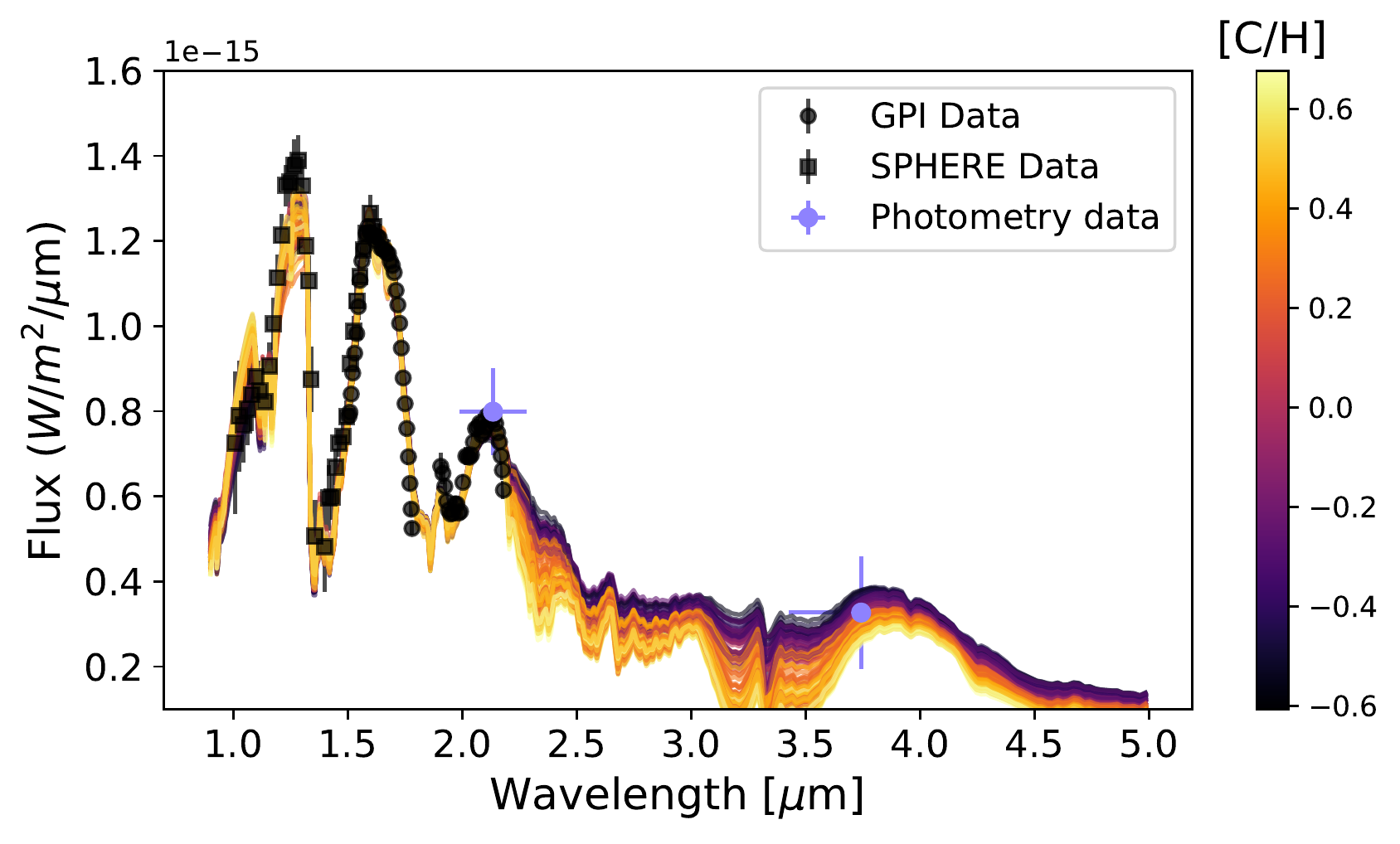}
    \caption{Random models drawn from the posterior of the baseline LRS retrieval (MgSiO$_3$, am), plotted over a larger wavelength range and color-coded by [C/H], the metallicity. There is a gradient in [C/H] in the $L$ ($\approx3.4-4.2~\mu$m) and $M$ ($\approx4.55-4.8~\mu$m) bands, which can be distinguished with comparable S/N LRS in these bands. The GPI and SPHERE data are shown in black, and we also overplot the photometric data points from \citet{crepp_TRENDS_2016}, which are not included in the retrievals but nonetheless agree with our models.}
    \label{fig:lowres_long_wave}
\end{figure}

We note that \citet{molliere_Retrieving_2020} were able to obtain much better constraints on the composition of HR~8799e, which also has a cloudy atmosphere, using LRS data sets from 0.95-2.5~$\mu$m. Their LRS had SNR between 4-11 per wavelength bin, much lower than the SNR of our LRS (between 20-60 per wavelength bin). Unlike \citet{molliere_Retrieving_2020}, however, our study does not have LRS in the second half of $K$ band ($2.2-2.5~\mu$m), which contains a strong CO bandhead as well as significant H$_2$O and CH$_4$ opacities. When we compute the CO abundances from our LRS retrievals, we find that they are not well constrained, with 1$\sigma$ intervals that are $\gtrsim3$ wider than the CO constraint from HRS. In Fig.~\ref{fig:lowres_long_wave}, we plot random draws of our baseline model color-coded by metallicity out to 5~$\mu$m. As shown, the models diverge quickly in the $2.2-2.5~\mu$m range. The fact that we miss this crucial wavelength region could explain why \citet{molliere_Retrieving_2020} obtain more robust constraints on atmospheric abundances, and P-T profiles that agree better with self-consistent models than we do, despite using data with a lower SNR.

Fig.~\ref{fig:lowres_long_wave} also shows a clear gradient in metallicity beyond 2.5~$\mu$m. In some of our cloudy LRS retrievals, we see a covariance between metallicity and cloud mass fraction, where lower metallicities correspond to higher cloud mass fractions, as well as larger, more physically consistent radii (see Fig.~\ref{fig:lowres_corner}). The degeneracy between metallicity and cloud mass fraction might arise because both molecular opacities and clouds contribute opacity, and our data has insufficient wavelength coverage to probe more regions where the gas and cloud opacities are sufficiently different. From the LRS retrievals, we consistently find a factor of $\sim2-3$ more CH$_4$ and H$_2$O than observed in the HRS, implying that the LRS retrievals could be compensating for our imperfect cloud models by increasing the gas opacities. 

Using a more flexible cloud model might alleviate some of these issues. For example, \citet{burningham_Cloud_2021} retrieved the 1-15~$\mu$m LRS of a field L dwarf and found the data preferred silicate clouds much higher up than the predicted cloud base locations from condensation curves. In addition, their retrieved cloud particles also have smaller sizes (sub-micron) than predicted by the EddySed model (a few microns). Similarly, \citet{luna_Empirically_2021} found that sub-micron cloud particles at lower pressures than predicted by EddySed are required to fit the mid-IR silicate feature ($\approx8-10~\mu$m) of many L dwarfs. They found that the microphysical cloud model CARMA \citep{turco_OneDimensional_1979, toon_Multidimensional_1988, gao_Sedimentation_2018} allows them to fit their data much better and even place constraints on which cloud species are producing the observed absorption feature.

Both the above-mentioned studies benefited from data at $\sim10~\mu$m that significantly help with constraining cloud properties. Thus, to obtain better abundance measurements with LRS, it is not only important to obtain full coverage in the near-IR (which we lack), but also to acquire data in the mid-IR. JWST can obtain low- and medium-resolution spectroscopy of brown dwarfs spanning the near- to mid-IR wavelengths using the NIRSpec and MIRI instruments. Future ground-based instruments such as SCALES at Keck \citep{stelter_Update_2020} will also provide LRS in the mid-IR. 

\section{Joint retrievals} \label{sec:joint_results}

In this section, we describe the results of joint retrievals to both the HRS and LRS for HD~4747~B. In practice, we set up two radiative transfer routines with \texttt{petitRADTRANS} using line-by-line (for HRS) and correlated-k (for LRS) opacity sampling respectively. The HRS and LRS models share the same atmospheric parameters and priors, but each has some unique parameters (e.g. RV and $v\sin{i}$ for HRS, GP kernel parameters for LRS). Within one nested sampling retrieval, we add the log likelihoods from the HRS and LRS components at each step of sampling to get the total log likelihood. We consider both clear and cloudy EddySed models for our joint retrievals. 

\begin{figure}[t!]
    \centering
    \includegraphics[width=\linewidth]{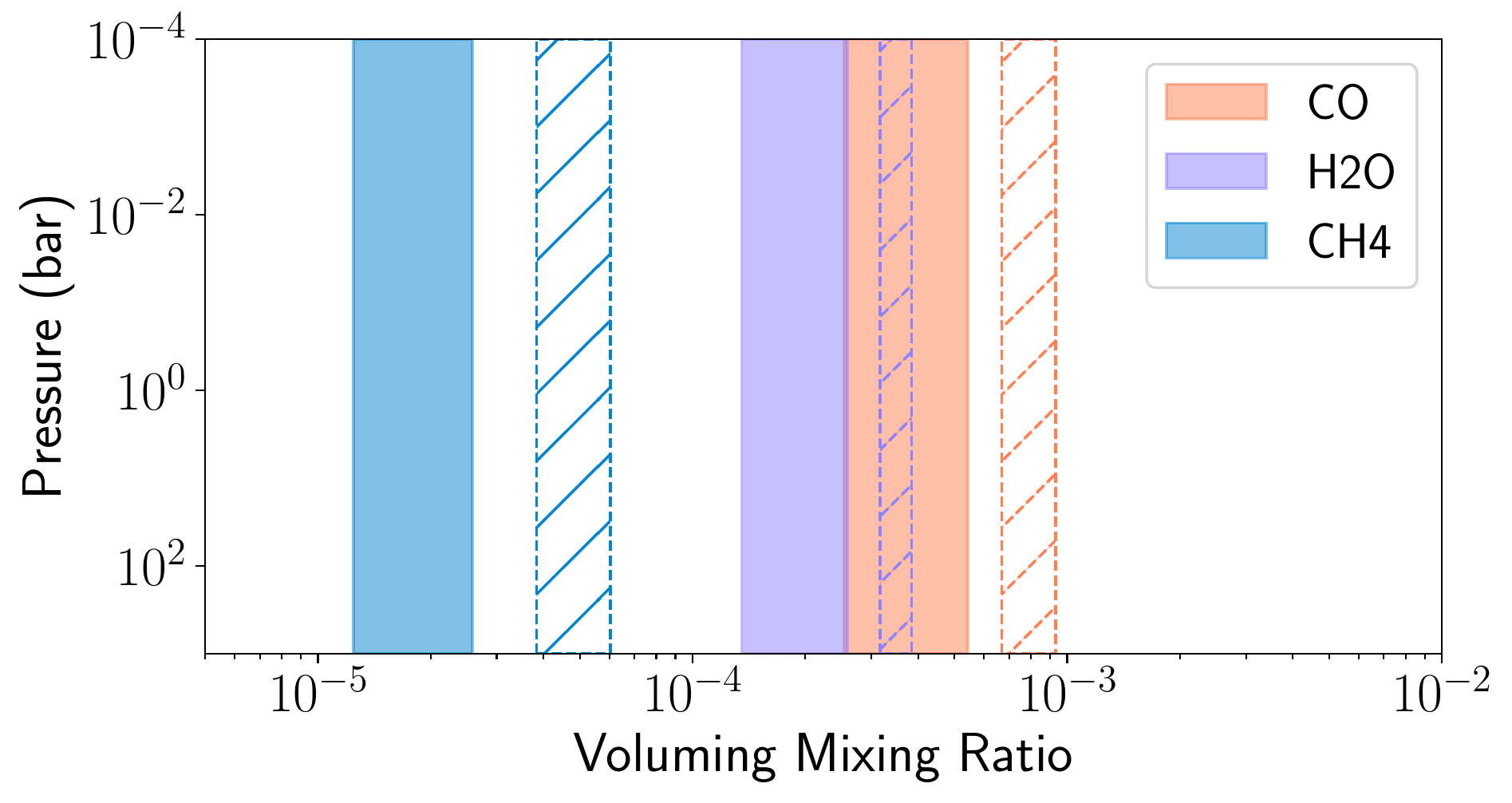}
    \caption{Filled areas: $1\sigma$ intervals for the CO, H$_2$O, and CH$_4$ abundances from our KPIC HRS retrieval. Hatched areas: the same for a joint retrieval (HRS + LRS). The retrieved abundances are 2-3 times higher in the joint retrieval, while the relative abundance ratios between species stays roughly the same (which produces a similar C/O). This highlights the fact that HRS is better at constraining relative abundances than absolute abundances. }
    \label{fig:abunds_joint_v_high}
\end{figure}

Because the LRS prefer clouds, the cloudy model (MgSiO$_3$, am) is overwhelmingly preferred in our joint retrieval, with a Bayes factor in excess of $10^{34}$ compared to the clear model. From the cloudy model, we retrieve C/O$=0.70\pm0.03$ and [C/H]$=0.34\pm0.07$. The retrieved uncertainties on these parameters are lower than in the HRS-only retrieval (which had C/O$=0.66\pm0.04$ and [C/H]$=-0.10^{+0.18}_{-0.15}$). In addition, the C/O from our joint fit is consistent with the C/O from our HRS fit. This is not surprising, because the HRS places tight constraints on the relative line depths (and hence the relative abundance ratios) of CH$_4$, H$_2$O, and CO. However, the joint fit pushes the metallicity to higher values, which corresponds to increased gas abundances as shown in Fig.~\ref{fig:abunds_joint_v_high}. The joint fit results translate to a $>4\sigma$ discrepancy in [C/H] between HD~4747~A\&B, while there is no discrepancy if we take the results from the HRS fit. This implies that the joint fit might be compensating for inadequacies in modeling clouds by increasing the gas opacities, as discussed in \S~\ref{sec:shortcoming_lrs} for the LRS-only case. We ran additional joint retrievals where we varied the cloud parameters (e.g. adding Fe clouds) and found similar results.

If we compare the log likelihoods of the HRS part of the joint fit to that from the HRS-only fit, we find that the HRS is fit less well by $\approx e^{10}$ (which translates to $\approx4\sigma$) in the joint fit, implying a trade-off between fitting the LRS and HRS. We can qualitatively compare the LRS S/N per wavelength bin to the CCF S/N of the HRS, which approximates the total constraining power of the HRS. When including all molecules in our model, we find a CCF S/N of $\approx15$ for the HRS. For the LRS, the average S/N per wavelength bin is $\approx 20$ for the SPHERE $YJH$ data and $\approx60/30$ between the GPI $H/K_1$ data. This explains why the joint fits prioritize fitting the LRS at the expense of the HRS.

As discussed in \S~\ref{sec:lrs_results}, the LRS are very model-sensitive and additionally contaminated by correlated noise. For this reason, we adopt the HRS-only results as the best estimate of HD~4747~B's atmospheric properties in this paper (see first row of Table~\ref{table:bayes_factors}). We leave it to future work, preferably aided by longer wavelength coverage in LRS, to achieve a more satisfactory joint retrieval.

\section{Discussion} \label{sec:discuss}

\subsection{Next steps for high-resolution spectroscopy} \label{sec:hrs_advantage}
Our KPIC HRS provide a better handle on the atmospheric abundances of HD~4747~B, and are less sensitive to model choices than our LRS. In fact, our $K$ band HRS are essentially agnostic to clouds in the brown dwarf's atmosphere; all retrieved parameters are consistent independent of our chosen cloud model (Fig.~\ref{fig:kpic_corner}). As discussed in \S~\ref{sec:cloud_insensitive}, this is because our HRS covers a wavelength region ($2.29-2.49~\mu$m) of high molecular opacity, and probes emission across a wide range of atmospheric pressures where cloud opacity is negligible (up to $10^{-2}$ bars in line cores). While clouds affect the continuum near $1~\mu$m in the LRS, they have little effect on the line depths across the wavelength range of our HRS. The relative lines depths are sensitive to relative molecular abundances, which directly constrains C/O. These results advocate for using HRS to measure atmospheric abundances. 

In the future, it is important to explore whether these findings hold true for other substellar objects. In upcoming papers, we will present KPIC HRS retrievals of brown dwarf companions and giant planets spanning a range of effective temperatures and surface gravities. Ultimately, it would also be useful to constrain cloud properties with HRS. For transmission spectroscopy, \citet{gandhi_Seeing_2020} found that their simulated near-IR HRS for warm Neptunes are more sensitive to molecular abundances than LRS for the same reasons highlighted in this study. While both clouds and metallicity affect the line depths in HRS, \citet{gandhi_Seeing_2020} show that increasing the wavelength coverage (e.g. going from 0.9-1.7~$\mu$m to 0.9-2.5~$\mu$m) helps distinguish between clouds and metallicity and provide better constraints on both. Thus, if we wish to obtain constraints on clouds and abundances at the same time, it would be important to extend our current HRS to a broader range of wavelengths. KPIC Phase II will allow us to obtain $L$ band data ($\approx3.4-4.1~\mu$m) to complement existing $K$ band data \citep{delorme_Keck_2021a}, and future upgrades could benefit from including $H$ and $J$ bands as well. 

In this study, we have assumed that the atmosphere of HD 4747~B is globally uniform. However, it would be important to examine the impact of 3-D effects, including non-uniform cloud coverage. Past studies with photometry or LRS show that many brown dwarfs exhibit clear rotational variability signals \citep[e.g.][]{Apai_HST_2013, zhou_Cloud_2018, biller_simultaneous_2018, manjavacas_Revealing_2021, vos_Let_2022}, which appear to be caused by inhomogeneities in their cloud properties. Therefore, time-resolved observations are important to understanding clouds and 3-D effects. 

With HRS, we can use the time-varying line depth and shape to map the 2-D brightness distributions of these objects \citep[e.g.][]{crossfield_global_2014}. In this paper, we used 1 hour of KPIC data for HD~4747~B. Given our measured $v\sin{i}$ and assuming a radius of $0.8\Rj$, we would expect a 5 or 7 hour rotation period if $i$ is equal to the orbital inclination or $i=90\degree$. Thus, it may be possible to sample a full rotation period within a single observing night, with the caveat that measurements of the true rotation period remain difficult for high-contrast companions \citep[][]{biller_highcontrast_2021}.

\subsection{Methane and the presence of disequilibrium chemistry} \label{sec:discuss_diseq_chem}
Our HRS retrievals indicate that the ratio of CO/CH4 (VMR) is $\approx10$ times higher than expected by equilibrium chemistry (see \S~\ref{sec:diseq_chem}). To gain more physical intuition, we convert the quench pressure from our HRS retrievals to an estimate of the vertical diffusion coefficient, $K_{\rm zz}$. To do this, we match the chemical timescale of the CO-CH$_4$ reaction from \citet{Zahnle_methane_2014} with the mixing timescale $\tau_{\rm mix} = L^2 / K_{\rm zz}$. While the length scale $L$ is typically taken to be the pressure scale height $H$ for lack of a better estimate, \citet{smith_Estimation_1998} show that this assumption is not valid across several reactions in the atmospheres of Jupiter and Neptune. In fact, \citet{smith_Estimation_1998} find that $L\approx0.1H$, which changes the inferred $K_{\rm zz}$ by two orders of magnitude. Similarly,  \citet{ackerman_Precipitating_2001} also note that the mixing length is generally shorter than the pressure scale height $H$ in stable atmospheric regions. Due to the uncertainty in $L$, we adopt  $L=\alpha H$, where $\alpha$ is a scaling factor, and $H= \frac{k_{B} T}{\mu m g}$ ($\mu$: mean molecular weight, $g$: surface gravity, $T$: the local temperature). For each value of quench pressure from our posteriors, we compute the necessary quantities to derive a posterior for $K_{\rm zz}$. For instance, if $\alpha=0.1$, we find $K_{\rm zz} = 5\times10^{8} - 1\times10^{12}~\rm{cm^2 s^{-1}}$ ($1\sigma$). On the other hand, if $\alpha=1$, we obtain $K_{\rm zz} = 5\times10^{10} - 1\times10^{14}~\rm{cm^2 s^{-1}}$.  

There have been few quantitative measurements of $K_{\rm zz}$ for substellar companions. \citet{miles_Observations_2020a} used M-band LRS to constrain the CO abundance and estimate the vertical diffusion coefficient, $K_{\rm zz}$, for seven field brown dwarfs. However, their objects have $T_{\rm eff}$ between 250-750~K, much colder than HD~4747~B. In terms of objects with $T_{\rm eff} \gtrsim 1000~$K, \citet{barman_simultaneous_2015} reported a detection of CH$_4$ in HR~8799b ($T_{\rm eff}\sim1000~$K) with Keck/OSIRIS data, which they used to estimate $K_{\rm zz}$ between $10^6-10^8~\rm{cm^2 s^{-1}}$. However, this CH$_4$ detection was not confirmed by an independent study \citep{petitditdelaroche_Molecule_2018}, and recently \citet{ruffio_Deep_2021} concluded that future higher-resolution follow up is needed to resolve the discrepant CH$_4$ signal strengths found by different analyzes. \citet{ruffio_Deep_2021} point out that if the CH$_4$ abundance was over-estimated by \citet{barman_simultaneous_2015}, that would imply a larger $K_{\rm zz}$. Using LRS, \citet{molliere_Retrieving_2020} found a well-constrained quench pressure for HR~8799e ($T_{\rm eff}\sim1100~$K) from \texttt{petitRADTRANS} retrievals, which could similarly be converted to a $K_{\rm zz}$ constraint. In summary, our finding HD~4747~B, which is $\sim300-400~$K hotter than HR~8799b\&e and much older (a few Gyr from \S~\ref{sec:host}) than most directly imaged planets, represents an important new data point because hotter objects are expected to be closer to equilibrium, making chemical disequilibrium processes harder to detect \citep[e.g.][]{moses_Chemical_2013}.

\citet{Zahnle_methane_2014} provide an upper limit for $K_{\rm zz}$ from mixing length theory \citep{Gierasch_convect1985} assuming full convection. For HD~4747~B, their Equation 4 translates to an upper limit of $\approx10^9~\rm{cm^2 s^{-1}}$. Depending on $\alpha$, our retrieved $K_{\rm zz}$ either exceeds this upper limit by $\gtrsim 2\sigma$ (if $L=H$), or is very close to this limit (if $L=0.1H$). Together, this suggests that convection is driving the vertical mixing in HD~4747~B, and that the mixing efficiency is likely close to its predicted maximum. We check whether our inferred $K_{\rm zz}$ makes sense by comparing them to those predicted by self-consistent atmospheric models with disequilibrium chemistry from \citet{karalidi_Sonora_2021} and Mukherjee et al. (2022, in prep). For an object with properties similar to HD~4747~B, our measured CH$_4$ VMR is consistent with $K_{\rm zz}\sim10^{8}-10^{12}$ in these models (with the assumption that $L=H$). These values of $K_{\rm zz}$ are roughly consistent with our estimate based on $P_{\rm quench}$, and also near the upper limit from \citet{Zahnle_methane_2014}. On the modeling front, it would be valuable to carry out 3-D hydrodynamical simulations \citep[e.g.][]{zhang_Globalmean_2018,tan_Atmospheric_2021a} of brown dwarf interiors to independently estimate $K_{\rm zz}$ \citep{tan_Jet_2022} and compare the results to that inferred by our data. Such simulations could also reveal what physical processes might cause a discrepancy between mixing length theory and our observations. 

\subsection{Dynamical versus Spectroscopic Mass Constraints} \label{sec:mass_prior}
For a majority of substellar companions observed by direct imaging, there are no dynamical mass constraints. To assess whether our mass prior plays an important role in the results, we repeat our HRS and LRS retrievals with the baseline cloud model but use uniform priors in mass from 10 to 100 \Mj (`free-mass'). For the HRS free-mass retrieval, we find that all parameters change by less than $1\sigma$ compared to the mass-prior retrieval. The mass itself shows a broad distribution (33-76~\Mj at $1\sigma$) that encompasses the dynamical mass. Because our KPIC HRS are not flux calibrated, the radius is not well constrained. In this case, we get large uncertainties in the spectroscopic mass because mass is inferred from the retrieved surface gravity, which depends on the poorly-constrained radius.

Our LRS free-mass retrieval also yields posteriors for all parameters that are consistent between 1-2~$\sigma$ with the mass-prior retrieval. Furthermore, the mass retrieved by the LRS is $59^{+7}_{-8}~\Mj$, which agrees within about $1\sigma$ with the dynamical mass. This provides confidence that reasonable mass constraints can be placed on substellar objects from LRS. The radius retrieved is $0.77\pm0.03~\Rj$, consistent with evolutionary model predictions and close to the radius from the mass-prior retrieval, suggesting the two retrievals find a similar surface gravity. 

\subsection{Atmospheric abundances of HD~4747~AB} \label{sec:discuss_abunds}
We retrieve [C/H] and [O/H] values that are $1\sigma$ consistent with those of the host star, as discussed in \S~\ref{sec:overview_abund}. Both companion and the star are mildly sub-solar in terms of their metal content. However, our retrieved C/O=$0.66\pm0.04$ is higher by approximately $2\sigma$ than the stellar C/O=$0.48\pm0.08$. 

The question is whether the marginal discrepancy in C/O is from astrophysical or systematic reasons. 
For example, \citet{wang_Retrieving_2022} carried out retrieval experiments on simulated HRS ($2.2-2.35~\mu$m, $R=35,000$) and found that their formal error bars are likely under-estimated due to systematic errors at the $\sim0.15$ level in C/O. Using KPIC HRS from 2.23-2.33~$\mu$m, they found $\approx1-1.5\sigma$ discrepancies between the [C/H] and [O/H] abundances of HR~7672~A and B, another benchmark brown dwarf system. On an earlier study of benchmark brown dwarfs, \citet{line_Uniform_2015} quoted $1\sigma$ uncertainties of $0.2-0.3$ in their brown dwarf C/O (much larger than our formal C/O uncertainty of $0.04$), and concluded that a $2\sigma$ agreement between the stellar and companion C/O is sufficiently good given the caveats. It is also possible that the uncertainties on stellar abundances are under-estimated given non-LTE effects \citep{line_Uniform_2015}. 

Another factor that might contribute to the $2\sigma$ discrepancy in C/O is uncertainties in the chemistry of condensates. The chemical model of \texttt{petitRADTRANS} we use accounts for the equilibrium condensation of various species and reports the global (rather than gas phase) C and O abundances \citep{molliere_petitRADTRANS_2019}. In particular, species such as MgSiO$_3$ and Mg$_2$SiO$_4$ contain 3 or 4 oxygen atoms per molecule, and are expected to hold a significant portion of O \citep{line_Uniform_2015}. From our HRS retrievals, we find that $\approx18\%$ of O is condensed into solids such as MgSiO$_3$. In order to decrease the global C/O of the brown dwarf by $\approx0.1$ (therefore making the companion and stellar C/O agree at the $1\sigma$ level), we require a $\sim20\%$ increase in the net O abundance. Keeping everything else unchanged, this means the MgSiO$_3$ mass fraction, which is predicted by the chemical model to be $\sim2\times10^{-3}$ in our retrievals, needs to be doubled to $\sim4\times10^{-3}$. From the LRS retrievals, the cloud base MgSiO3 fraction can be as high as $10^{-2}$. Therefore, a factor of $\sim2$ uncertainty in the abundance of MgSiO$_3$ could make our C/O consistent at the $1\sigma$ level with the stellar value. 

Given these caveats, we conclude that the $2\sigma$ difference between our retrieved C/O for HD~4747~B and the stellar value is not significant, and HD~4747~AB are consistent with being chemically homogeneous. Chemical homogeneity is expected by models where brown dwarf companions form via gravitational fragmentation in molecular clouds \citep[e.g.][]{padoan_Mysterious_2004} or massive protostellar disks \citep[e.g.][]{stamatellos_Brown_2007}. Simulations suggest that brown dwarfs typically form as part of unstable, high-order multiple systems, which undergo chaotic interactions that reduce the multiplicity over time \citep[e.g.][]{bate_formation_2002, thies_Tidally_2010, bate_stellar_2012}. With a semi-major axis of 10 au, HD~4747~B is unlikely to have been directly affected by such encounters, but its relatively high orbital eccentricity ($\approx0.73$) could encode such a dynamically `hot' past. 

\section{Conclusions} \label{sec:conclude}
Using high-resolution spectra ($R\sim35,000$) obtained by Keck/KPIC, we retrieve [C/H]=$-0.10^{+0.18}_{-0.15}$, [O/H]=$-0.18^{+0.18}_{-0.15}$, and C/O=$0.66\pm0.04$ for the benchmark brown dwarf companion HD~4747~B (formal error bars). The C and O abundances are consistent with the stellar values to $\lesssim1\sigma$, while the C/O ratio is consistent at the $2\sigma$ level, as expected for a binary-star like formation scenario. This shows that we can measure the atmospheric abundances for high contrast substellar companions to the $20\%$ level with KPIC and our current modeling framework, which \citet{wang_Retrieving_2022} also show for another benchmark brown dwarf. We outline some other key findings from our study below.

We measure precise abundances from the KPIC HRS ($2.29-2.49~\mu$m), which are insensitive to our choice of cloud model. Our abundance measurements suggest that HD~4747~B has a CO/CH$_4$ ratio that is 10 times higher than predicted by equilibrium chemistry, corresponding to a quench pressure of $50-260$ bars ($1\sigma$). This translates to a high vertical diffusion coefficient $K_{\rm zz}$ which depends on the assumed length scale $L$. However, even if $L$ is ten times smaller than the pressure scale height, we get $K_{\rm zz}=5\times10^{8} - 1\times10^{12}~\rm{cm^2 s^{-1}}$, which implies a mixing strength that is at or above the upper limit predicted by mixing length theory. 

The composition retrieved from our LRS (1-2.2~$\mu$m) are both sensitive to model choices, and can be biased by the presence of speckles. For this reason, HRS provides a more reliable picture of the atmospheric composition in the current data sets, although the LRS could be improved with additional observation at longer wavelengths including the $L$ and $M$ bands. Despite these challenges, the current LRS does provide a spectroscopic mass estimate that is $1\sigma$ consistent with the dynamical mass for the brown dwarf. 

Although our joint retrieval results are likely biased by the limited LRS wavelength coverage, joint analyzes of LRS and HRS remain a promising avenue to constrain cloud properties and abundances simultaneously and provide a more complete picture of substellar atmospheres. When extended wavelength coverage is available, it would also be important to consider possible 3-D effects, including patchy clouds. These might be constrained by obtaining multiple spectra sampling a rotation period. Additional modeling work on condensation, chemistry, and vertical mixing rates are also important to inform future observational results.

\acknowledgements
We thank the referee for helpful comments that improved the manuscript. J.X. thanks Michael Zhang for advice in computing the CH$_4$ opacities, Jonathan Fortney for discussions on chemical disequilibrium, and Konstantin Batygin for discussions on brown dwarf formation. 

We wish to recognize and acknowledge the very significant cultural role and reverence that the summit of Maunakea has always had within the indigenous Hawaiian community. We are most fortunate to have the opportunity to conduct observations from this mountain. This research has benefitted from the SpeX Prism Library maintained by Adam Burgasser at http://www.browndwarfs.org/spexprism. Funding for KPIC has been provided by the California Institute of Technology, the Jet Propulsion Laboratory, the Heising-Simons Foundation, the Simons Foundation, and the United States National Science Foundation Grant No. AST-1611623. AV acknowledges funding from the European Research Council (ERC) under the European Union's Horizon 2020 research and innovation programme, grant agreements No. 757561 (HiRISE). HAK acknowledges support from the President's and Director's Research \& Development Fund Program, which is jointly funded by the Jet Propulsion Laboratory and the California Institute of Technology under a contract with the National Aeronautics and Space Administration. The computations presented here were conducted in the Resnick High Performance Center, a facility supported by Resnick Sustainability Institute at the California Institute of Technology.

\facilities{Keck (KPIC)}
\software{\texttt{petitRADTRANS}~\citep{molliere_petitRADTRANS_2019}, \texttt{dynesty}~\citep{speagle_DYNESTY_2020}, \texttt{PyAstronomy}~(\url{https://github.com/sczesla/PyAstronomy})}

\bibliography{Imaging.bib}{}
\bibliographystyle{aasjournal}

\appendix
\section{Orbit fits for HD~4747~B} 
\begin{figure*}
\centering
\begin{subfigure}
  \centering
  \includegraphics[width=.4\linewidth]{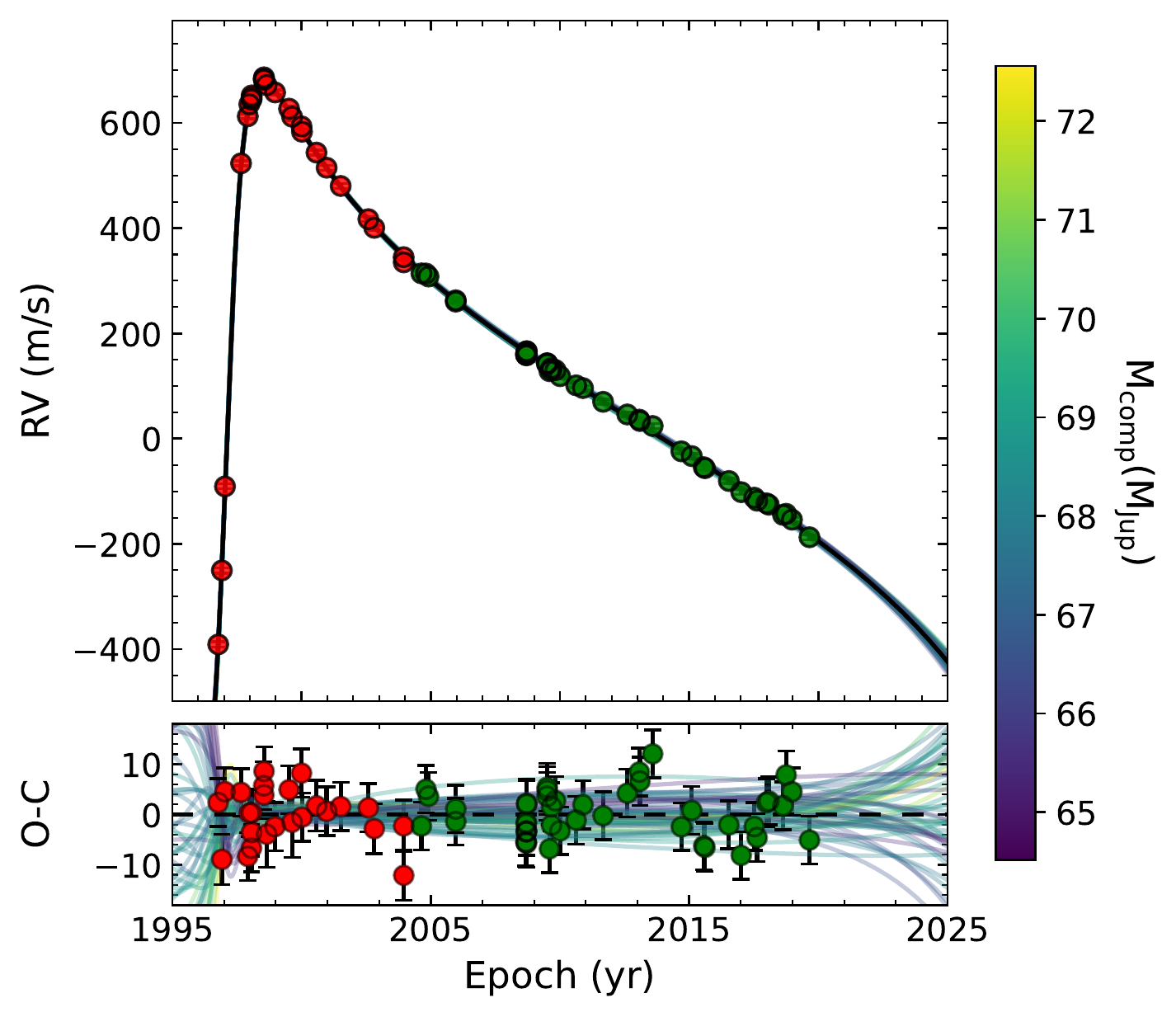}
  \hspace{10mm}
\end{subfigure}
\begin{subfigure}
  \centering
  \includegraphics[width=.4\linewidth]{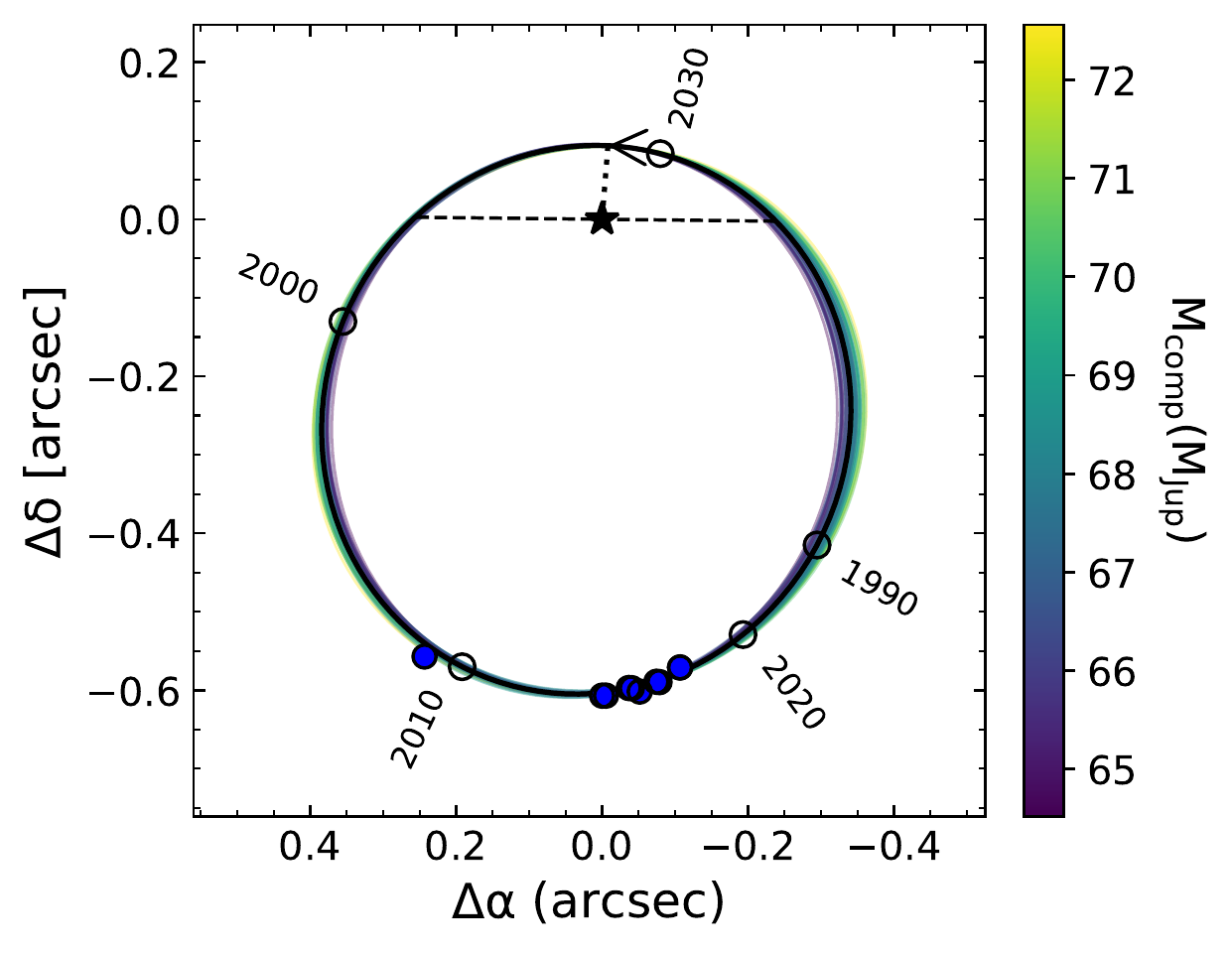}
\end{subfigure}
\begin{subfigure}
  \centering
  \includegraphics[width=.6\linewidth]{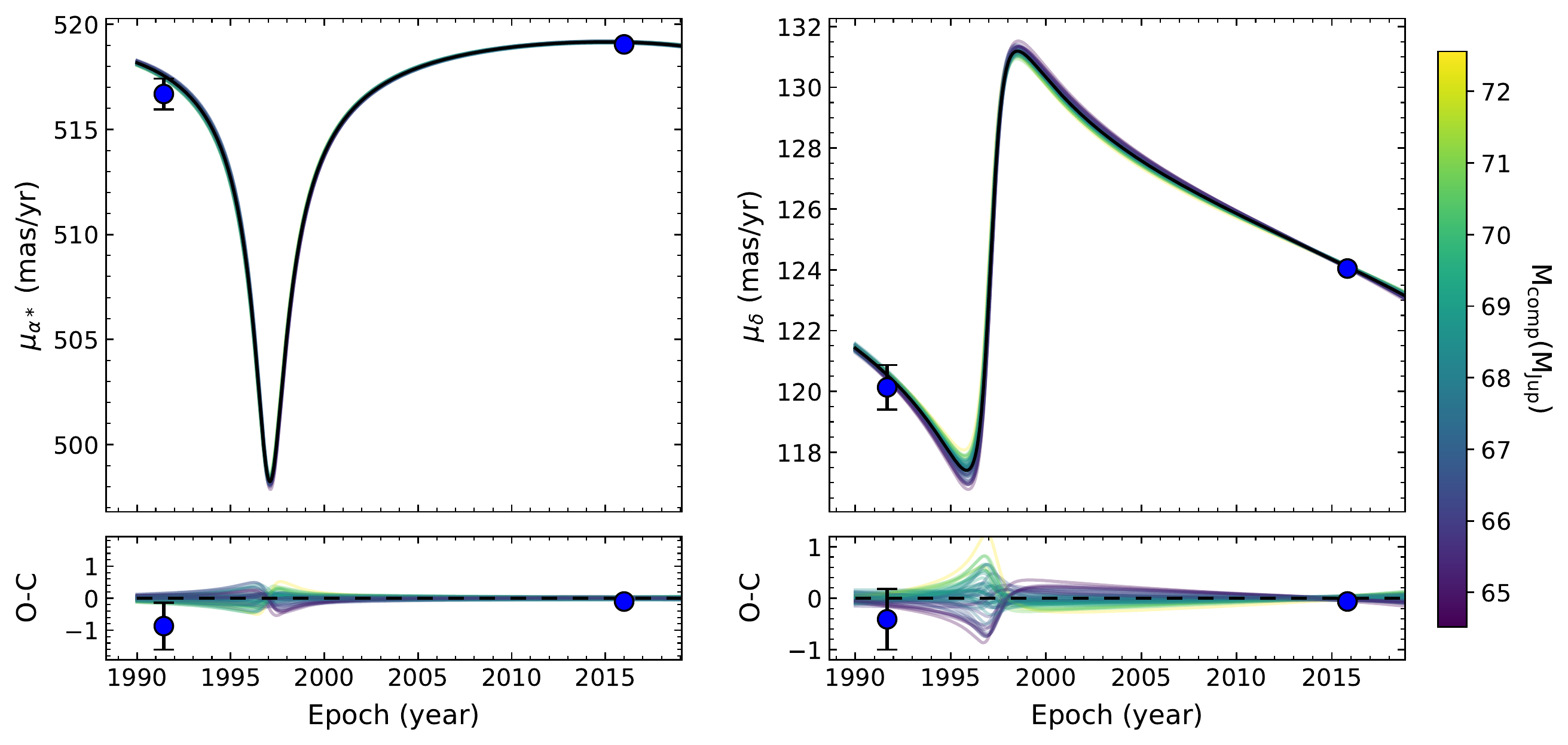}
\end{subfigure}
\caption{Results from a joint fit to host star radial velocity (top left), relative astrometry (top right), and absolute astrometry (bottom panel) for the HD~4747 system. The data together constrain the orbital parameters and mass of both the companion and host star well (Table~\ref{tab:orbit}).}
\label{fig:orbit}
\end{figure*}

Our orbit fit for the HD~4747 system is shown in Fig.~\ref{fig:orbit}.

\section{Extracted Low-resolution spectrum and GPI astrometry}
Our extracted spectrum for HD~4747~B based on observations with GPI \citep{crepp_GPI_2018} and SPHERE \citep{peretti_orbital_2019} are given in Table~\ref{tab:gpi_spec}. Our relative astrometry measurements based on the GPI data are listed in Table~\ref{tab:gpi_astrom}.

\startlongtable
\begin{deluxetable}{lcc}
\tablecaption{Extracted Low-Resolution Spectrum for HD~4747~B} \label{tab:gpi_spec}
\tablehead{\colhead{Wavelength ($\mu$m)} & \colhead{Flux ($10^{-15}\times$ $W/\rm{m}^2/\mu$m)} & \colhead{Flux error ($10^{-15}\times$ $W/\rm{m}^2/\mu$m)}} 
\startdata
\hline
SPHERE ($YJH$) \\
\hline
1.008 & 0.726 & 0.167 \\ 
1.026 & 0.789 & 0.131 \\ 
1.044 & 0.767 & 0.088 \\ 
1.063 & 0.806 & 0.085 \\ 
1.081 & 0.839 & 0.086 \\ 
1.098 & 0.88 & 0.073 \\ 
1.116 & 0.848 & 0.071 \\ 
1.138 & 0.823 & 0.059 \\ 
1.159 & 0.907 & 0.056 \\ 
1.176 & 1.006 & 0.061 \\ 
1.194 & 1.114 & 0.054 \\ 
1.212 & 1.213 & 0.05 \\ 
1.23 & 1.33 & 0.049 \\ 
1.248 & 1.338 & 0.06 \\ 
1.267 & 1.378 & 0.061 \\ 
1.285 & 1.389 & 0.06 \\ 
1.302 & 1.33 & 0.063 \\ 
1.318 & 1.188 & 0.051 \\ 
1.33 & 1.106 & 0.061 \\ 
1.34 & 0.875 & 0.077 \\ 
1.358 & 0.506 & 0.086 \\ 
1.399 & 0.482 & 0.106 \\ 
1.418 & 0.596 & 0.078 \\ 
1.432 & 0.598 & 0.053 \\ 
1.447 & 0.668 & 0.042 \\ 
1.463 & 0.725 & 0.033 \\ 
1.479 & 0.741 & 0.034 \\ 
1.495 & 0.788 & 0.038 \\ 
1.511 & 0.912 & 0.036 \\ 
1.526 & 0.987 & 0.036 \\ 
1.54 & 1.059 & 0.038 \\ 
1.553 & 1.117 & 0.038 \\ 
1.568 & 1.179 & 0.04 \\ 
1.582 & 1.217 & 0.04 \\ 
1.599 & 1.266 & 0.044 \\ 
1.613 & 1.235 & 0.041 \\ 
\hline
GPI ($H$) \\
\hline
1.506 & 0.789 & 0.022 \\ 
1.51 & 0.798 & 0.022 \\ 
1.516 & 0.841 & 0.024 \\ 
1.522 & 0.89 & 0.025 \\ 
1.531 & 0.936 & 0.026 \\ 
1.539 & 0.983 & 0.028 \\ 
1.547 & 1.046 & 0.031 \\ 
1.554 & 1.107 & 0.031 \\ 
1.562 & 1.153 & 0.032 \\ 
1.572 & 1.183 & 0.033 \\ 
1.581 & 1.218 & 0.033 \\ 
1.589 & 1.233 & 0.034 \\ 
1.597 & 1.22 & 0.033 \\ 
1.605 & 1.216 & 0.032 \\ 
1.613 & 1.21 & 0.031 \\ 
1.621 & 1.201 & 0.032 \\ 
1.63 & 1.21 & 0.033 \\ 
1.638 & 1.208 & 0.032 \\ 
1.646 & 1.183 & 0.031 \\ 
1.654 & 1.181 & 0.032 \\ 
1.662 & 1.18 & 0.032 \\ 
1.67 & 1.172 & 0.033 \\ 
1.678 & 1.171 & 0.033 \\ 
1.686 & 1.153 & 0.032 \\ 
1.695 & 1.143 & 0.031 \\ 
1.703 & 1.126 & 0.03 \\ 
1.711 & 1.084 & 0.029 \\ 
1.719 & 1.05 & 0.028 \\ 
1.727 & 1.007 & 0.027 \\ 
1.735 & 0.949 & 0.025 \\ 
1.743 & 0.878 & 0.023 \\ 
1.751 & 0.818 & 0.022 \\ 
1.758 & 0.759 & 0.021 \\ 
1.765 & 0.693 & 0.019 \\ 
1.772 & 0.63 & 0.017 \\ 
1.777 & 0.57 & 0.018 \\ 
1.781 & 0.525 & 0.015 \\ 
\hline
GPI ($K_1$) \\
\hline
1.892 & 0.547 & 0.063 \\ 
1.898 & 0.627 & 0.054 \\ 
1.905 & 0.608 & 0.081 \\ 
1.907 & 0.671 & 0.032 \\ 
1.916 & 0.654 & 0.026 \\ 
1.924 & 0.624 & 0.02 \\ 
1.932 & 0.588 & 0.019 \\ 
1.941 & 0.567 & 0.019 \\ 
1.95 & 0.559 & 0.018 \\ 
1.96 & 0.561 & 0.019 \\ 
1.969 & 0.582 & 0.02 \\ 
1.977 & 0.582 & 0.018 \\ 
1.985 & 0.561 & 0.018 \\ 
1.993 & 0.564 & 0.018 \\ 
2.003 & 0.633 & 0.018 \\ 
2.016 & 0.694 & 0.021 \\ 
2.025 & 0.693 & 0.022 \\ 
2.033 & 0.692 & 0.022 \\ 
2.041 & 0.697 & 0.022 \\ 
2.049 & 0.727 & 0.022 \\ 
2.059 & 0.759 & 0.023 \\ 
2.069 & 0.762 & 0.025 \\ 
2.077 & 0.772 & 0.027 \\ 
2.086 & 0.744 & 0.022 \\ 
2.094 & 0.752 & 0.023 \\ 
2.103 & 0.764 & 0.028 \\ 
2.111 & 0.79 & 0.024 \\ 
2.12 & 0.79 & 0.028 \\ 
2.129 & 0.778 & 0.028 \\ 
2.138 & 0.787 & 0.024 \\ 
2.147 & 0.771 & 0.025 \\ 
2.155 & 0.749 & 0.022 \\ 
2.163 & 0.728 & 0.021 \\ 
2.17 & 0.695 & 0.022 \\ 
2.176 & 0.662 & 0.023 \\ 
2.181 & 0.615 & 0.022 \\ 
2.183 & 0.471 & 0.025 \\ 
\enddata
\tablecomments{This table is available in its entirety in machine-readable form.}
\end{deluxetable}

\begin{deluxetable}{lcc}
\tablecaption{Extracted GPI Astrometry for HD~4747~B} \label{tab:gpi_astrom}
\tablehead{\colhead{Time (BJD)} & \colhead{Separation (arcsec)} & \colhead{Position angle (deg)}} 
\startdata
2457380.5 & $0.5989\pm0.002$ & $183.9\pm0.2$ \\
2457381.5 & $0.5984\pm0.002$ & $183.5\pm0.2$ \\
\enddata
\end{deluxetable}

\section{Priors and posteriors for retrieval parameters}
Here we list the priors on our retrieved parameters and include joint posterior distributions of selected parameters from our baseline HRS and LRS retrievals.

\begin{deluxetable*}{ll|ll}[t!]
\tablecaption{Priors of the HD~4747~B retrieval. $\mathcal{U}$ stands for a uniform distribution, with two numbers representing the lower and upper boundaries. $\mathcal{G}$ stands for a Gaussian distribution, with numbers representing the median and standard deviation. (a) and (b): These priors follow \citet{molliere_Retrieving_2020}. $P_{\rm phot}$ is the pressure where $\tau=1$, and $T_{\rm connect}$ is the uppermost temperature of the `photospheric' layer, and is computed by setting $\tau=0.1$ in the Eddington Approximation (see eq. 1 and 2 in \citealt{molliere_Retrieving_2020}). This prior, along with those on $T_1$ and $T_2$ are used to prevent temperature inversions. (c) $\tilde{X}_{\rm MgSiO_3/Fe}$ represents the scaling factor for the cloud mass fraction, so that $\rm{log} (\tilde{X}_{\rm MgSiO_3/Fe})=0$ refers to a fraction equal to the equilibrium mass fraction. $f_{\rm sed}$, $K_{\rm zz}$, and $\sigma_{\rm g}$ are parameters in the EddySed cloud model \citep{ackerman_Precipitating_2001}. When fitting molecular abundances directly (e.g. in \S~\ref{sec:ch4_detect}), we use the same mass fraction prior on all molecules included.}
\label{tab:samp_prior}
\tablehead{Parameter & Prior & Parameter & Prior}
\startdata
Mass ($\Mj$) & $\mathcal{G}(67.2, 1.8)$ & $\rm C/O$ & $\mathcal{U}(0.1,1.6)$ \\
Radius ($\Rj$) & $\mathcal{U}(0.6, 1.2)$ & $\rm [Fe/H]$ & $\mathcal{U}(-1.5,1.5)$ \\
$T_1$ ($K$) & $\mathcal{U}(0 , T_2)$ & ${\rm log}(P_{\rm quench}/{\rm bar})$ & $\mathcal{U}(-4, 3)$ \\
$T_2$ ($K$) & $\mathcal{U}(0 , T_3)$ & $f_{\rm sed}$ & $\mathcal{U}(0, 10)$ \\
$T_3$ ($K$) & $\mathcal{U}(0 , T_{\rm connect})^{\rm (a)}$ & ${\rm log}(K_{\rm zz}/\rm{cm^2 s^{-1}})$ & $\mathcal{U}(5, 13)$ \\
$T_{\rm int}$ ($K$) & $\mathcal{U}(700, 2500)$ & $\sigma_{\rm g}$ & $\mathcal{U}(1.05, 3)$ \\
$\alpha$ & $\mathcal{U}(1,2)$ & ${\rm log}(\tilde{X}_{\rm MgSiO_3})^{\rm (c)}$ & $\mathcal{U}(-2.3, 1)$ \\
${\rm log}(\delta)$ & $P_{\rm phot} \in [10^{-3}, 100]^{\rm (b)}$ & ${\rm log}(\tilde{X}_{\rm Fe})$ & $\mathcal{U}(-2.3, 1)$ \\ \hline
Additional parameters for HRS \\ \hline
RV (km/s) & $\mathcal{U}(-30 , 30)$ & $v\sin{i}$ (km/s) & $\mathcal{U}(0, 50)$ \\ 
Error multiple & $\mathcal{U}(1 , 4)$ & Flux scale (counts) & $\mathcal{U}(0, 200)$ \\ \hline
Gaussian process parameters for LRS \\ \hline
${\rm log}(f_{\rm amp})$ & $\mathcal{U}(10^{-4}, 1)$ & ${\rm log}(l)$ ($\mu$m) & $\mathcal{U}(10^{-3}, 0.5)$ \\ \hline
Mass fraction of molecules \\ \hline
log(MMR) & $\mathcal{U}(10^{-1}, 10^{-7})$ & \phd & \phd \\
\enddata
\end{deluxetable*}

\begin{figure*}
\centering
\includegraphics[width=\linewidth]{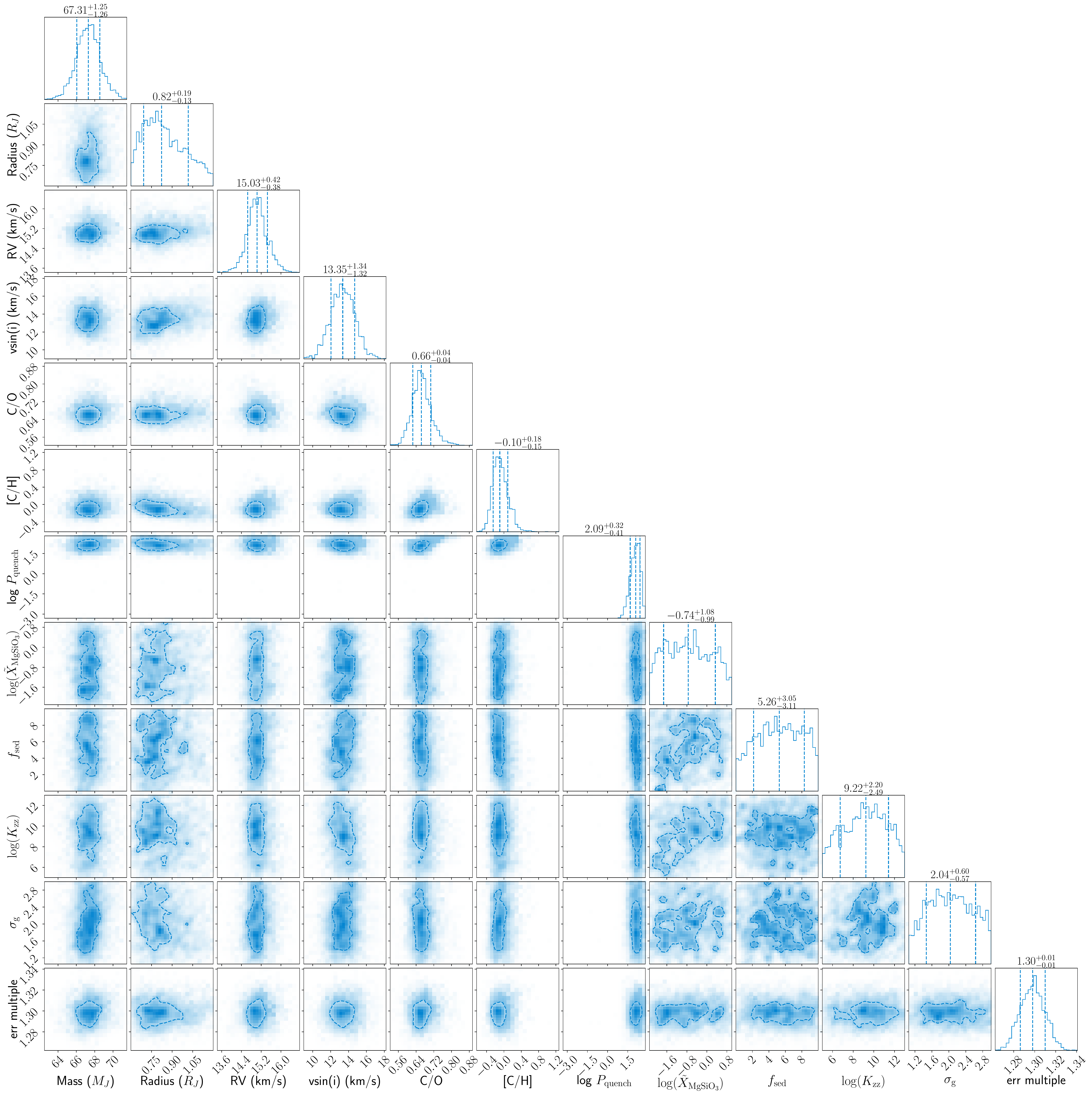}
\caption{Joint posterior distributions for the HRS retrieval of HD~4747~B. We omit the P-T profile parameters, which are better visualized by the P-T plot in Fig.~\ref{fig:pt_profile}.}
\label{fig:corner_hrs}
\end{figure*}

\begin{figure*}
\centering
\includegraphics[width=\linewidth]{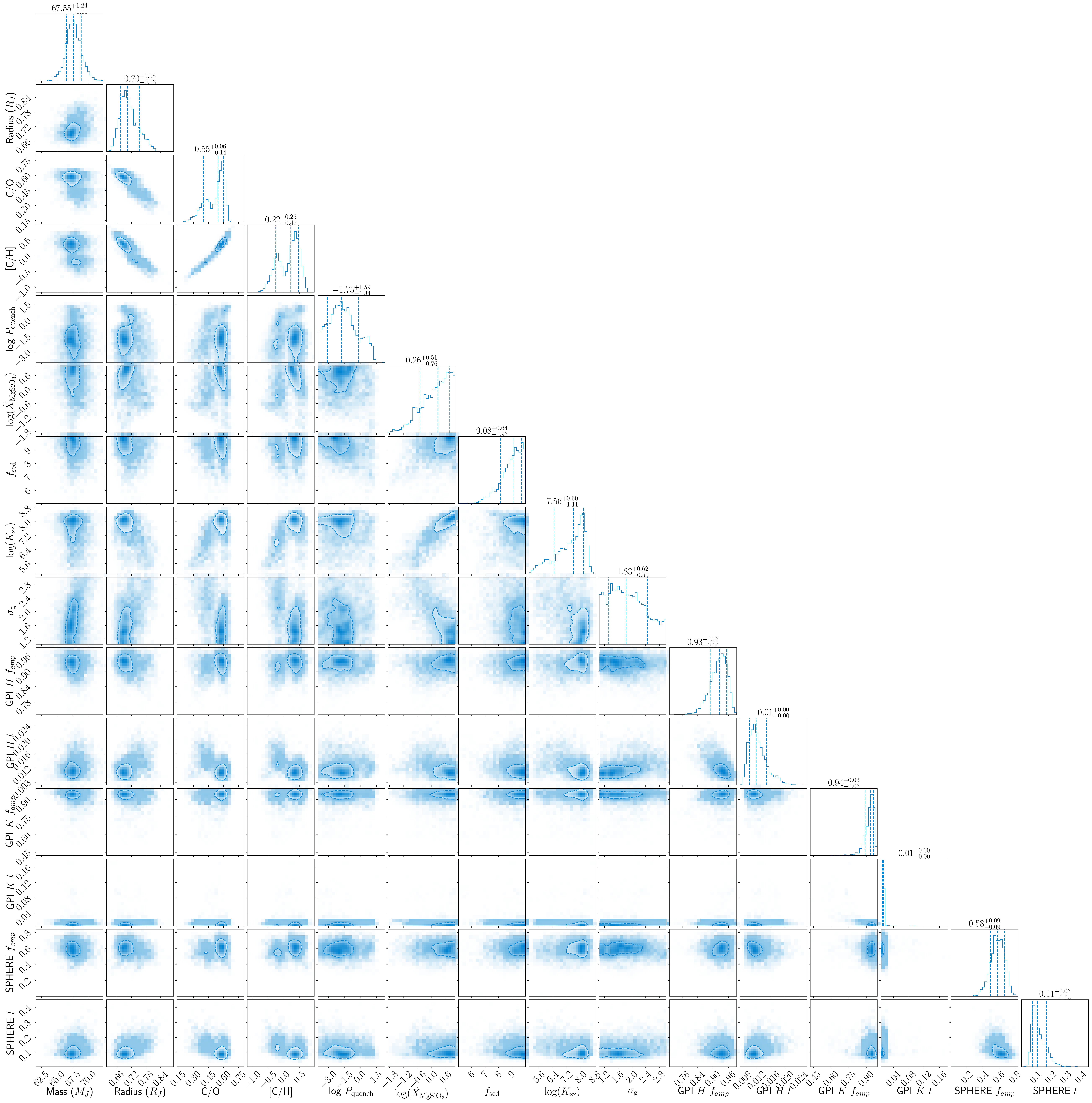}
\caption{Joint posterior distributions for the LRS retrieval of HD~4747~B. We omit the P-T profile parameters, which are better visualized by the P-T plot in Fig.~\ref{fig:pt_profile}. The distribution for a few parameters are bi-modal.}
\label{fig:corner_lrs}
\end{figure*}

\end{document}